\documentclass[aps,nofootinbib,superscriptaddress]{revtex4}

\usepackage{amsfonts,amssymb,amsthm,amsmath}

\usepackage{color,psfrag}
\usepackage[dvips]{graphicx}

\newcommand{\N}{{\mathbb N}}
\newcommand{\R}{{\mathbb R}}

\newcommand{\cA}{{\mathcal A}}
\newcommand{\cB}{{\mathcal B}}
\newcommand{\cE}{{\mathcal E}}
\newcommand{\cF}{{\mathcal F}}
\newcommand{\cG}{{\mathcal G}}

\newcommand{\cR}{{\mathcal R}}
\newcommand{\cO}{{\mathcal O}}

\newcommand{\cD}{{\mathcal D}}
\newcommand{\cC}{{\mathcal C}}

\newcommand{\cU}{{\mathcal U}}
\newcommand{\SU}{\mathrm{SU}}

\newcommand{\SL}{\mathrm{SL}}
\newcommand{\SO}{\mathrm{SO}}
\newcommand{\U}{\mathrm{U}}

\newcommand{\be}{\begin{equation}}
\newcommand{\ee}{\end{equation}}
\newcommand{\beq}{\begin{eqnarray}}
\newcommand{\eeq}{\end{eqnarray}}
\newcommand{\bes}{\begin{eqnarray}}
\newcommand{\ees}{\end{eqnarray}}

\newcommand{\tabl} [2] {\begin{array} {#1} #2 \end{array}}

\newcommand{\su}{{\mathfrak su}}

\renewcommand{\u}{{\mathfrak u}}
\renewcommand{\sl}{{\mathfrak sl}}
\newcommand{\so}{{\mathfrak so}}
\newcommand{\g}{{\mathfrak g}}
\newcommand{\la}{\langle}
\newcommand{\ra}{\rangle}

\newcommand{\f}{\frac}

\def\fE{{\mathfrak E}}

\def\nn{\nonumber}

\def\cF{{\cal F}}

\def\mone{^{{-1}}}
\def\cop{\Delta}

\newcommand{\id}{\mathbb{I}}

\def\dr{\rightarrow}
\def\ie{\textit{i.e. \hspace{.05mm}}}

\def\UQ{{\cU_{q}(\su(2))}}
\def\ot{\otimes}
\def\one{{\bf 1}}
\def\act{\triangleright}

\def\bt{{\bf t}}
\def\btt{\tilde{\bf t}}

\def\tjm{{\bf t}^{j}_{m}}

\def\tjma{{\bf t}^{j_{1}}_{m_{1}}}

\def\jm{|jm\ra}

\def\demi{\f{1}{2}}
\def\UUQn{{\cU_{q}(\u(n))}}

\def\com#1{[ #1 ]}

\newcommand{\CG} [2] {\,_{q}\textbf{C}\begin{array} {#1} #2 \end{array}}
\def\CGa{\CG{c@{}c@{}c}{j_1& j_2& j\\m_1 & m_2 & m}}
\newcommand{\mCG} [2] {\,_{q\mone}\textbf{C}\begin{array} {#1} #2 \end{array}}
\def\qCG{$q$-Clebsch-Gordan }
\def\qsj{ q-$6j$-symbol }
\newcommand{\qsixj} [2] {\, \left\lbrace \begin{array} {#1} #2 \end{array}\right\rbrace \, }
\def\psir{\psi_{\cR}}

\newtheorem{theorem}{Theorem}[section]
\newtheorem{lemma}[theorem]{Lemma}

\newtheorem{definition}[theorem]{Definition}

\newcommand\T{\rule{0pt}{4ex}}       
\newcommand\B{\rule[-4ex]{0pt}{0pt}} 

\begin{document}

\title{Observables in Loop Quantum Gravity with a cosmological constant
 }

\author{{\bf Ma\"it\'e Dupuis}}\email{maite.dupuis@gravity.fau.de}
\affiliation{University Erlangen-Nuremberg, Institute for Theoretical Physics III, Erlangen, Germany}

\author{{\bf Florian Girelli}}\email{fgirelli@uwaterloo.ca}
\affiliation{Department of Applied Mathematics, University of Waterloo, Waterloo, Ontario, Canada}
\affiliation{University Erlangen-Nuremberg, Institute for Theoretical Physics III, Erlangen, Germany}


\begin{abstract}
An open issue in loop quantum gravity (LQG) is the introduction of a non-vanishing cosmological constant $\Lambda$. In 3d, Chern-Simons theory provides some guiding lines:  $\Lambda$ appears in the quantum deformation of the gauge group. The Turaev-Viro model, which is an example of spin foam model is also defined in terms of a quantum group. By extension, it is believed that in 4d, a quantum group structure could encode the presence of $\Lambda\neq0$. \\
In this article, we introduce by hand the quantum group $\UQ$ into the LQG framework, that is we deal with $\UQ$-spin networks. We explore some of the consequences, focusing in particular on the structure of the observables.  Our fundamental tools are tensor operators for $\UQ$. We review their properties and give an explicit realization of the spinorial and vectorial ones. We construct the generalization of the $\U(N)$ formalism  in this deformed case, which is given by the quantum group $\UUQn$. We are then able to build geometrical observables, such as the length, area or angle operators ... We show that these operators characterize a quantum discrete hyperbolic geometry in the 3d LQG case. Our results confirm that the use of quantum group in LQG can be a tool to introduce a non-zero cosmological constant into the theory.
\end{abstract}
\maketitle


\tableofcontents


\section*{Introduction}

%
\paragraph*{\bf Background:}
There are different proposals to understand the nature of the cosmological constant $\Lambda$. It can be interpreted as encoding some type of vacuum energy (see \cite{stefano1, stefano2, sola} and references therein) or as a coupling constant just like the Newton's constant $G$. The loop quantum gravity and spinfoam frameworks use the latter interpretation which is motivated by the   seminal works of Witten \cite{witten-3d}, and later of Fock and Rosly \cite{fock}, and Alekseev, Grosse, Schomerus \cite{Alek1, Alek2}.  Indeed, in a 3d space-time, one can rewrite General Relativity with a (possibly zero) cosmological constant as a Chern-Simons gauge theory\footnote{This is actually an extension of General Relativity since  degenerated metrics are allowed.}. The general phase space structure of the theory  for any metric signature and sign of $\Lambda$ can be treated in a nice unified way \cite{catherine}, using Poisson-Lie groups \cite{chari}, the classical counterparts of quantum groups. The quantization procedure leads \textit{explicitly} to a quantum group structure. The full construction, from phase space to quantum group is usually called \textit{combinatorial quantization} \cite{fock, Alek1, Alek2}. \\
We can also quantize 3d gravity using the spinfoam approach. In this approach, 3d gravity is formulated as a BF theory. When $\Lambda=0$, this is the well-known Ponzano-Regge model (both Euclidian or Lorentzian), based on the irreducible unitary representations of the relevant gauge group.  When $\Lambda\neq0$,  the quantum group structure is introduced by hand. The Ponzano-Regge model is deformed, using irreducible unitary representations of the relevant quantum deformation of the gauge group. This is then called the Turaev-Viro model \cite{viro}. The argument consolidating the incorporation of the cosmological constant into a spinfoam model through a quantum group comes from the semi-classical limit. Indeed, the asymptotics of the deformed $\{6j\}_q$ symbol, entering into the definition of the Turaev-Viro model, goes to the Regge action with a cosmological constant in the regime  $\ell_p\ll \ell \ll R$. \\ 
The third approach to quantize gravity is the canonical approach, \ie the loop quantum gravity approach (LQG). In this case, performing the classical hamiltonian analysis to General Relativity, the cosmological constant only appears in the Hamiltonian constraint. This means that the kinematical space is the same whether $\Lambda=0$ or not. In particular this kinematical space (where the Gauss constraint has been solved) is based on the classical relevant gauge group.

\smallskip

Therefore at this stage, quantum groups naturally appear only in the combinatorial quantization of Chern-Simons. Different quantum groups are revealed according to the metric signature and the sign of the cosmological constant.  When $\Lambda\neq 0$, we obtain $q$-deformed version of the gauge group $\cU_q(\mathfrak{g})$, where $\mathfrak{g}$ is the Lie algebra of the gauge group $G=\SL(2,\R)$ in the Lorentzian case, $\SU(2)$ in the Euclidean case, with $q$ function of the Planck scale and the cosmological radius $R=\sqrt{|\Lambda|}$. The deformation parameter $q$ can be real or complex.  A nice way to recall what is $q$ according to the sign of $\Lambda$ and the signature is to consider $q=exp\left({-\frac{\hbar G \sqrt{\Lambda}}{\sqrt{c^2}}}\right)$ and posing $c^2>0$ in the Lorentzian case and   $c^2<0$ in the Euclidian case \cite{bernd}. Note that this trick gives $q$ or $q^{-1}$. 
The full relevant quantum group arising from the combinatorial quantization is  $\cD(\cU_q(\mathfrak{g}))$, the Drinfeld double of $\cU_q(\mathfrak{g})$. When $\Lambda=0$, we get the Drinfeld double $\cD(\cU(\mathfrak{g}))$ with a non-commutative parameter given by $\kappa = \ell_p$ in  units $\hbar=1=c$. A list of the different quantum groups relevant for 3d gravity  is given in the first table below.

\medskip

 Since classically, the Chern-Simons formulation and the standard formulation of General Relativity are equivalent (modulo the degenerated metrics), we can wonder whether the Chern-Simons combinatorial quantization formalism, LQG and the spinfoam framework are related in some ways. It can be shown explicitly in the Euclidian case, with $\Lambda>0$, that the Chern-Simons quantum model and  the  Turaev-Viro model are related, more precisely,  the Turaev Viro amplitude is the square of the Chern-Simons amplitude \cite{CS-TV}. 
 On the other hand,  it seems difficult to relate the LQG formalism, when  $\Lambda \neq 0$,  to a spin foam model based on a quantum group if we assume that the LQG kinematical space is based on a classical group such as $\SU(2)$. 
 
When $\Lambda=0$, it is also possible to relate the Chern-Simons amplitude and the Ponzano Regge amplitude \cite{louapre},  which allows to identify a hidden symmetry given by the Drinfeld double $\cD(\cU(\mathfrak{g}))$ in the Ponzano-Regge model. 
Still when $\Lambda=0$,  explicit links between  LQG  and the spinfoam framework \cite{ale-karim} or between the Chern Simons combinatorial quantization and  LQG \cite{karim-cat} have been identified.  Note also that we can identify a hidden quantum group structure (the Drinfeld double) in LQG when $\Lambda=0$ \cite{louapre, karim, karim-cat}, which is consistent  with the other approaches.    The different cases for 3d gravity are summarized in the first table. For more details, we refer to the excellent review \cite{bernd}.

%

\begin{center}
\begin{tabular}{|c|c|c|c|} \hline \label{tablQG}
 Signature & $\Lambda$ & Quantum group  & QG models \\ \hline
 \begin{tabular}{cc}    \begin{tabular}{c}  \hspace{2mm}\\Euclidian \\\hspace{2mm}\end{tabular} \\ \hline   \begin{tabular}{c}\hspace{2mm} \\ Lorentzian \\\hspace{2mm}\end{tabular}  \end{tabular} &  \begin{tabular}{cc}$\Lambda>0$  \\$\Lambda=0$  \\$\Lambda<0$    \\ \hline$\Lambda>0$  \\$\Lambda=0$  \\$\Lambda<0$\end{tabular}
&
 \begin{tabular}{cc}$\cD(\UQ)$, $q= e^{i \f{\ell_p}{R}}$  \\ $\cD(\cU(\su(2)))$, $\kappa = \ell_p$  \\ $\cD(\UQ)$, $q= e^{\f{\ell_p}{R}}$    \\ \hline
 $\cD(\cU_q(\sl(2,\R)))$, $q=e^{-\f{\ell_p}{R}}$ \\
  $\cD(\cU(\sl(2,\R)))$, $\kappa = \ell_p$  \\
$\cD(\cU_q(\sl(2,\R)))$, $q=e^{-i\f{\ell_p}{R}}$
 \end{tabular} &
  \begin{tabular}{cc}Chern-Simons $\stackrel{\text{\cite{CS-TV}} }{{\leftrightarrow}}$ Turaev-Viro $\stackrel{?}{\leftrightarrow}$ LQG    \\ Chern-Simons $\stackrel{\text{\cite{louapre}}}{\leftrightarrow}$ Ponzano-Regge$\stackrel{\text{\cite{ale-karim}}}{\leftrightarrow}$ LQG $\stackrel{\text{\cite{karim-cat}}}{\leftrightarrow}$ Chern-Simons \\ Chern-Simons $\stackrel{?}{\leftrightarrow}$ Turaev-Viro $\stackrel{?}{\leftrightarrow}$ LQG   \\ \hline
Chern-Simons $\stackrel{?}{\leftrightarrow}$ Turaev-Viro $\stackrel{?}{\leftrightarrow}$ LQG  \\Chern-Simons $\stackrel{\text{\cite{louapre}}}{\leftrightarrow}$ Ponzano-Regge$\stackrel{\text{\cite{ale-karim}}}{\leftrightarrow}$ LQG $\stackrel{\text{\cite{karim-cat}}}{\leftrightarrow}$ Chern-Simons   \\Chern-Simons $\stackrel{?}{\leftrightarrow}$ Turaev-Viro $\stackrel{?}{\leftrightarrow}$ LQG\end{tabular} \\\hline
 \end{tabular}
\end{center}


When dealing with 4d space-time, there is no Chern-Simons theory to guide us. Hence, it is postulated that the cosmological constant should also be introduced through a quantum group structure. From the spinfoam  approach, one then considers the model one prefers (Barrett-Crane (BC) or EPRL-FK) when $\Lambda=0$, based on the irreducible unitary representations of the gauge group and one deforms it \cite{yetter, roche, winston, muxin0}. To argue a posteriori, that this is the right thing to do, we can look at the asymptotic of the spinfoam amplitude and check we recover the Regge action with a cosmological constant \cite{limits}. 
It is quite interesting that the current "physical" EPRL spinfoam model defined in the Lorentzian case, with $\Lambda>0$ leads to a finite amplitude \cite{winston, muxin0}.

\medskip 

In 4d, we are not able to connect the Hamiltonian constraint arising in LQG to a spinfoam model, even when $\Lambda=0$. Just as in 3d, it is not clear at all why a quantum group structure should appear in the LQG framework. There exist few arguments to justify this postulate \cite{pranzetti}. We include now a table summarizing the different quantum group models appearing in 4d quantum gravity. 

\begin{center}
\begin{tabular}{|c|c|c|c|}\hline
 Signature & $\Lambda$ & Quantum group  & QG models \\ \hline
 \begin{tabular}{cc}   \begin{tabular}{c} \hspace{2mm}\\Euclidian \\\hspace{2mm}\end{tabular} \\ \hline   \begin{tabular}{c}\hspace{2mm} \\ Lorentzian \\\hspace{2mm}\end{tabular}  \end{tabular} &  \begin{tabular}{cc}$\Lambda>0$  \\$\Lambda=0$  \\$\Lambda<0$    \\ \hline$\Lambda>0$  \\$\Lambda=0$  \\$\Lambda<0$\end{tabular}
&
 \begin{tabular}{cc}$\cU_q(\so(4))$, $q= e^{i2\pi {\ell^2_p}{\Lambda}}$  \\ ?  \\ $\cU_q(\so(4))$, $q= e^{i2\pi \f{\ell^2_p}{\Lambda}}$    \\ \hline
 $\cU_q(\so(3,1))$, $q=e^{ {\ell^2_p}{\Lambda}}$  \\?  \\$\cU_q(\so(3,1))$, $q=e^{ \f{\ell^2_p}{\Lambda}}$  \end{tabular} &
  \begin{tabular}{cc} BC or EPRL-FK $\stackrel{?}{\Leftrightarrow}$ LQG    \\BC or EPRL-FK  $\stackrel{?}{\Leftrightarrow}$  LQG \\  BC or EPRL-FK  $\stackrel{?}{\Leftrightarrow}$  LQG  \\ \hline
BC or EPRL-FK  $\stackrel{?}{\Leftrightarrow}$ LQG  \\ BC or EPRL-FK  $\stackrel{?}{\Leftrightarrow}$ LQG  \\BC or EPRL-FK  $\stackrel{?}{\Leftrightarrow}$ LQG\end{tabular} \\\hline
 \end{tabular}
\end{center}


Several remarks can be made at this stage. The partition function of the  Plebanski action is invariant under the transformation $\Lambda \dr -\Lambda$ \cite{de pietri freidel}, which explains why we have the same quantum group for the different signs of the cosmological constant. This change of sign for $\Lambda$ is equivalent to $q\dr q\mone$. 

In the "physical" case (Lorentzian, $\Lambda>0$) in the EPRL-FK model,   spin networks encoding the quantum state  of space are defined in terms of $\UQ$, with $q$ \textit{real} \cite{winston, muxin0}.

We also emphasize \textit{en passant}, that the quantum deformation of the Lorentz group (in 3d or 4d) for $q$ complex are not understood.  

\medskip

\paragraph*{\bf Motivations:}
A common feature of the 3d and 4d quantum gravity is that it is hard to understand why a  $q$-deformation of the gauge group would appear  from the LQG perspective. 
Since we do not know how to solve the Hamiltonian constraint (for $\Lambda \neq 0$) and since we would like to compare the LQG approach with the well-known models coming from combinatorial quantization formalism and spinfoam, we would like to define LQG with a $q$-deformed group and see what the consequences are. We hope then to identify some hints pointing to the quantum group apparition in this context. In particular, if LQG defined in terms of a quantum group describes well quantum curved geometries, then this is a good sign that this could be a useful theory to consider.

To this aim, we need to understand the structure of the observables associated to spin networks defined using the representations of a quantum group. Not much work has been done in this context:  LQG with a quantum group has only been explored using the loop variables by Major and Smolin \cite{major1, major2, major3}.

When $\Lambda=0$, the structure of the observables for a spin network (or an intertwinner) is well understood, thanks to the spinor approach to LQG \cite{un0, un1, un2}. In particular it is possible to construct a closed algebra (a $\u(n)$ Lie algebra, where $n$ is the number of intertwinner legs) that generates all the observables acting on an intertwinner. This approach not only gives some information about the observable structure but it has been applied to different contexts, with many interesting results \cite{un0, un1, un2}. 
This formalism has  helped to understand that spin networks can be seen as the quantization of classical discrete geometries, the so called \textit{twisted geometries} \cite{twisted1, twisted2}. It allowed the construction  of a  new  Hamiltonian constraint in 3d Euclidian gravity \cite{valentin-etera}, such that the kernel of this constraint is given by the $6j$ symbol, \ie the Ponzano-Regge amplitude. It has provided the tools to implemented in a rigorous way the simplicity constraints, using the Gupta-Bleuler method,  to build a spinfoam model for Euclidian gravity ($\Lambda=0$) \cite{maite1}. 

Generalizing the spinor formalism to the quantum group case will help to better understand  the quantum gravity regime with a nonzero cosmological constant. Indeed, within this formalism, we should be able to construct an Hamiltonian constraint relating Turaev-Viro and LQG \cite{hamiltonian}, and we should be able to understand what is the relevant phase space for LQG, the space of curved twisted geometries \cite{phase space}.

\medskip

\paragraph*{\bf Main results:}

This generalization of the spinor formalism to the quantum group case is the main result of this paper. We have focused on the quantum group $\UQ$ with  $q$ real, which is therefore relevant for 3d Euclidian gravity with $\Lambda<0$ and the physical case, \ie 4d Lorentzian gravity with $\Lambda>0$. 

The key idea for this generalization is the use of \textit{tensor operators}. These are well-known in the quantum mechanical case for $\SU(2)$ \cite{sakurai}. Essentially, they are sets of operators that transform well under $\SU(2)$, \ie as a representation. They are  known in LQG under the name of \textit{grasping operators}. However they have not been studied intensively in this context. We  show  that considering these operators seriously naturally leads to the spinor approach to LQG. These tensor operators can be generalized to the quantum group case (more exactly they are defined for any quasi-triangular Hopf algebra) \cite{rittenberg}.

Given an $\UQ$ intertwinner with $n$ legs, we have identified  some sets of operators that transform well under $\UQ$. Due to the quantum group structure, they are much more complicated than their classical counterparts. In particular their commutation relations are pretty complicated. We have clarified  the   construction of  $\UQ$ intertwinner observables.   We show how there exists a fundamental algebra generating all observables, which is a deformation of the $\u(n)$ algebra.  We also discuss the geometric interpretation of some observables for 3d Euclidian LQG with $\Lambda<0$,  pinpointing the fact that the quantum group structure encodes as expected the notion of curved discrete geometry.  Some of these results were already announced in \cite{ours}.

\medskip

\paragraph*{\bf Outline of the paper:}
The paper is organized as follow. In section \ref{secOverview}, we recall the main features of $\UQ$, the $q$-deformed universal enveloping algebra of $\SU(2)$, with $q$ real. We recall as well the notion of $q$-harmonic oscillators which are used to build some  tensor operators explicit realizations.

Section \ref{secTO} is a  review about tensor operators for $\UQ$, the essential tools of our construction. Due to the nonlinearity of the quantum group structure, $\UQ$ tensor operators are more complicated than the standard $\SU(2)$ case.  In particular, due to the nontrivial nature of the quantum group action, the tensor product of tensor operators is highly nontrivial, which will make the construction of tensor operators acting on different legs of an intertwiner quite cumbersome, but necessary. 

Different explicit realizations of tensor operators for $\UQ$ are given in section \ref{realization}. 
 We recalled the results of Quesnes \cite{quesne} regarding spinor operators: their definition in terms of $q$-harmonic oscillators and their commutation relations for spinor operators acting on different legs. We have  extended this analysis to vector operators, which will be relevant for the construction of  the standard geometric operators.   

The main results of this paper are presented in section \ref{secResults} and \ref{lqg-result}. We discuss the general construction of observables for a $\UQ$ intertwiner. We construct a new realization of $\UUQn$ in terms of tensor operators, which is also invariant under the action of $\UQ$. We have identified the non-linear map relating  our invariant operators to the standard  $\UUQn$ Weyl-Cartan generators. We construct different geometric operators which we interpret in the context of 3d Euclidian LQG with $\Lambda<0$. We show how we get a quantization of the hyperbolic cosine law, a quantization of the length and of the area of a triangle. We pinpoint also how the presence of the cosmological constant allows for a notion of minimum angle. 

In the concluding section, we discuss the possible follow-ups of this tensor operator approach to LQG.  

We have also included some appendices to recall the definition of the hyperbolic cosine law as well as some  relevant formulae regarding the $\UQ$ recoupling coefficients.

\section{ $\UQ$ in a nutshell} \label{secOverview}

\subsection{Definition of $\UQ$}
In this section, we review the salient features of $\UQ$, which we shall extensively use, to fix the notations. 
We consider   $\UQ$, the $q$-deformation of the universal algebra of $\SU(2)$, with $q$ {\it real}, generated by $J_z, \; J_+, \; J_-$. We have the commutation relations
\be \label{commutationSU2q}
[ J_z, J_\pm]=\pm J_\pm, \quad [J_+, J_-]= [2J_z],  \textrm{ with }
[J_z]=\f{q^{J_z/2}-q^{-J_z/2}}{q^{1/2}-q^{-1/2}}.
\ee
For $qÊ\rightarrow 1$ the right-hand side of the second equation of \eqref{commutationSU2q} approaches $2J_z$ and we thus recover the usual Lie algebra $\su(2)$. 
$\UQ$ is equipped with a structure of  quasitriangular Hopf algebra $(\cop, \epsilon, S, \mathcal{R})$ \cite{chari, majid, kassel}.  
 \begin{itemize}
\item The   coproduct  $\Delta:\UQ\dr \UQ\otimes \UQ$ encodes physically  the total angular momentum of a 2-particle system.
\be \label{deform sum angular}
\Delta J_z=J_z \otimes \one + \one \otimes J_z, \quad \Delta J_\pm= J_\pm \otimes q^{J_z/2}+q^{-J_z/2}\otimes J_\pm.
\ee
 Considering the un-deformed case, we have 
\beq
(\cop J_\sigma)\, |j_1m_1, j_2m_2\ra = (J_\sigma \otimes \one + \one \otimes J_\sigma)  |j_1m_1j_2m_2\ra=  (J_\sigma ^{(1)} + J_\sigma^{(2)}) |j_1m_1j_2m_2\ra, \; \textrm{ where } \sigma = +,\,-,  \, z.
\eeq
In the deformed case, the addition of angular momenta \eqref{deform sum angular} is non-commutative, hence the addition of $q$-angular momenta depends on the order we set our particles. As we shall see, the braiding constructed using the $\mathcal{R}$-matrix  will allow to relate different orderings.
\item The counit $\epsilon:  \UQ \dr \UQ$ is defined such that $\epsilon(\id)=1, \, \epsilon(J_\sigma)=0$ for $\sigma=+, \, -, \, z$.
\item The antipode $S:\UQ\dr \UQ$ encodes in some sense the notion of inverse angular momentum.
\beq\label{antipode}
SJ_z=-J_z, \quad S J_\pm=-q^{\pm 1/2} J_\pm.
\eeq
\item The $\mathcal{R}$-matrix encodes the "amount" of non-commutativity of the coproduct, $\ie$  of the addition of angular momenta. Indeed, if we note $\psi: \UQ\ot \UQ \dr \UQ\ot \UQ$, the permutation, then we have that 
\beq\label{braiding}
(\psi\circ  \cop) X = \cR (\cop X) \cR\mone.
\eeq
In terms of the $\UQ$-generators, the $\mathcal{R}$-matrix can be written as
\beq \label{Rmatrix}
\mathcal{R}=\sum \cR_1\ot \cR_2 = q^{J_z\otimes J_z}\sum_{n=0}^\infty \f{(1-q^{-1})^n}{[n]!}q^{n(n-1)/4}(q^{J_z/2}J_+)^n\otimes (q^{-J_z/2}J_-)^n,
\eeq
where $[n]$ denotes the $q$-number $[n] \equiv \f{q^{\f{n}{2}}-q^{-\f{n}{2}}}{q^{\f{1}{2}}-q^{-\f{1}{2}}}$.
A co-commutative product would simply mean that $\cR=\one\ot\one$, which is obtained when $q\dr 1$ in \eqref{Rmatrix}. Further properties of the $\cR$-matrix are given in the Appendix \ref{rmat app}, in particular its expression in terms of Clebsch-Gordan coefficients. 

The non-co-commutativity of the coproduct implies that we have a "non-commutative" tensor product. Essentially, we would get a symmetric 2-particle system if the  permutation of the particles states does not affect the total observable, that is the permutation leaves invariant the coproduct,  $\psi\circ \cop=\cop$.

If it is non-co-commutative, as in the $\UQ$ case, we can still define a {\it deformed} permutation $\psi_{\cR}$ -- thanks to the existence of the $\cR$-matrix  \cite{rittenberg, kassel}.
\bes\label{deformedPerm}
\psir:  V\ot W &\dr& W\ot V \nn\\
v\ot w &\dr &\psir(|v, w\ra)\equiv \psi (\cR |v, w\ra) = \sum \psi (|\cR_{1} v ,\cR_{2} w\ra)  = \sum |\cR_{2} w , \cR_{1}v\ra.
\ees
Using the key property $(\psi \circ \cop) X = \cR (\cop X ) \cR\mone$, we have that
\beq\nn
\psir (X (|v, w\ra)) = \psi (\cR X( |v, w\ra))= \psi (\cR (\cop X) |v, w\ra)=  \psi ((\psi \circ \cop X) \cR |v, w\ra)= (\cop X) \psi (\cR |v, w\ra)= X(\psir(|v, w\ra)).
\eeq
Hence, the tensor product is only symmetric under this deformed notion of permutation.  From now on, we shall always consider this deformed permutation $\psir$ which is the natural notion of permutation in this quasi-triangular context.
\end{itemize}
The representation theory of $\UQ$ with $q$ real is very similar to the one of $\su(2)$ \cite{BiedenharnBook}. A representation $V^{j}$ is generated by the vectors $|j,m\ra$ with $j\in\N/2$ and $m\in \left\{ -j,..,j\right\}$.  The key-difference is that the action of the generators on these vectors generates  $q$-numbers. \bes\label{vector transformation}
&& J_{z}\, |jm\ra= m\; |jm\ra, \\
&& J_{\pm}\, |jm\ra = \sqrt{[ j \mp m][j\pm m +1]} \; |{j} {m\pm1}\ra.
\ees 
A Casimir operator can be defined as
\beq
C= J_{+}J_{-} + [J_{z}][J_{z}-1]= J_{-}J_{+} + [J_{z}][J_{z}+1].
\eeq
The tensor product of vectors $|j_{1}m_{1},  j_{2}m_{2}\ra$  can be decomposed into a linear combination of vectors using the $q$-Clebsh-Gordon (CG) coefficients $\CG{c@{}c@{}c} { j_{1}& j_{2} &j\\  m_1 & m_{2}& m}$.
\beq\label{TOrecoupling01}
|j_{1}m_{1}, j_{2}m_{2}\ra = {\sum_{j,m}} \,
 \CG{c@{}c@{}c}{j_1& j_2& j \\ m_1& m_2 & m} \jm, \quad j= |j_{1}-j_{2}|,..,j_{1}+j_{2}.
\eeq
Conversely, given a representation $V^{j}$ of $\UQ$ we can decompose it along two representations $V^{j_{1}}$ and $V^{j_{2}}$ of $\UQ$ (with $ |j_{1}-j_{2}|\leq j \leq j_{1}+j_{2}$ )
\be \label{TOrecoupling0}
\jm=\sum_{m_1, m_2} \, \CG{c@{}c@{}c}{j_1& j_2&j \\ m_1& m_2 & m} |j_{1}m_{1},j_{2}m_{2}\ra.
\ee
Acting with a generator $J_\sigma \, (\sigma=+,-,z)$ on the righthand side of  \eqref{TOrecoupling01} and with its coproduct on the lefthand side of \eqref{TOrecoupling01} we obtain a recursion relation for the CG coefficients  \cite{BiedenharnBook}. Such recursion relations can be taken as defining the CG coefficients.
\bes \label{CGrecursion}
&& J_z \, \act |j_1m_1,j_2m_2\ra = {\sum_{j,m}} \,  \CG{c@{}c@{}c}{j_1& j_2& j \\ m_1& m_2 & m} J_z \,\act  \jm \Leftrightarrow  \cop J_z\, |j_1m_1,j_2m_2\ra = {\sum_{j,m}} \,  \CG{c@{}c@{}c}{j_1& j_2& j \\ m_1& m_2 & m} J_z \, \jm   \nn \\
&& \qquad \qquad \qquad \qquad  \Rightarrow \; m_1+m_2=m \nn \\
&&   J_\pm\, \act  |j_1m_1,j_2m_2\ra = {\sum_{j,m}} \,  \CG{c@{}c@{}c}{j_1& j_2& j \\ m_1& m_2 & m} J_\pm \,\act  \jm  \Leftrightarrow  \cop J_\pm\, |j_1m_1,j_2m_2\ra = {\sum_{j,m}} \,  \CG{c@{}c@{}c}{j_1& j_2& j \\ m_1& m_2 & m} J_\pm \, \jm \nn \\
&& \quad \Rightarrow   \; q^{-\f{m_1}{2}} \left([j_2\pm m_2][j_2\mp m_2+1]\right)^\demi   \CG{c@{}c@{}c}{j_1& j_2& j \\ m_1& m_2\mp1 & m} + q^{\f{m_2}{2}} \left([j_1\pm m_1][j_1\mp m_1+1]\right)^\demi   \CG{c@{}c@{}c}{j_1& j_2& j \\ m_1\mp1& m_2 & m}   \nn\\
&&\qquad \qquad \qquad \qquad \qquad = \left([j\mp m][j\pm m+1]\right)^\demi \CG{c@{}c@{}c}{j_1& j_2& j \\ m_1& m_2 & m\pm 1}.
\ees
We refer to the Appendix \ref{CG app} for further  CG coefficients relevant properties. 

\medskip

Let us now introduce the notion of intertwiner for $\UQ$ which is a fundamental object in LQG. An intertwinner  is a vector $|\iota_{j_{1}..j_{N}}\ra= \sum_{m_{i}} c_{m_{1}..m_{N}} |j_{1}m_{1},.., j_{N}m_{N}\ra \in V^{j_{1}}\ot..\ot V^{j_{N}}$ which is invariant under the action of $\UQ$. 
\beq\label{invariant vector}
J_{\alpha} \act |\iota_{j_{1}..j_{N}}\ra = \left[(\one \ot ..\one \ot \cop )\circ.. \circ (\one \ot \cop)\circ \cop \right]( J_{\alpha}) |\iota_{j_{1}..j_{N}}\ra=0, \; \alpha=\pm,z.
\eeq 
 Note that since the coproduct is co-associative, we have no issue on how to compose the coproducts. In the case of $N=3$, \eqref{invariant vector} is equivalent to the  recursion relations which define the CG coefficients. A normalized 3-valent intertwiner is then uniquely defined by
\beq\label{trivalent}
|\iota_{j_{1}j_{2}j_{3}}\ra= \sum_{m_i} \f{(-1)^{j_3-m_3}q^{-\f{m_3}2}}{[2j_3+1]^\demi} \CG{c@{}c@{}c}{j_{1}& j_{2} & j_{3}\\m_{1}& m_{2}& -m_{3}} |j_{1}m_{1}, j_{2}m_{2},j_{3}m_{3} \ra. \nn
\eeq

\medskip
Another ingredient which we shall use extensively in the following sections, 
is the adjoint action of $\UQ$ on an operator $\cO$. It differs from  the usual adjoint action of $\su(2)$ given by a commutator. The $\UQ$ adjoint action of the generators $J_\sigma$ is explicitly given by 
\bes
J_z\act \cO= [J_z,\cO], \quad J_\pm\act \cO = J_\pm \cO q^{-J_z/2} - q^{\pm\demi}q^{-J_z/2}\cO J_\pm.
\ees
The following lemma is useful to relate quantities which are invariant under the adjoint action and the different Casimir one can construct. This is especially relevant in our case since the commutator and the adjoint action are not coinciding. 

\begin{lemma}\label{adjact-casimir}
Let $\cC\in \UQ$  invariant under the adjoint action, then $\cC$ commutes with the generators $J_{\sigma}$, $\sigma=+,-,z$. Conversely, if $\cC\in \UQ$ commutes with $J_{\sigma}$, then it is invariant under the adjoint action.
\end{lemma}

\subsection{$q$-harmonic oscillators and the Schwinger-Jordan trick}\label{SJ}

To account for the deformation, we consider a pair of $q$-harmonic oscillators,  comprising annihilation operators  $\alpha_{i} = a,b$, creation operators $\alpha^{\dagger}_{i} = a^{\dagger},b^{\dagger}$ and number operators $N_{\alpha_i}=N_a, N_b$, to construct representations of $\UQ$. There are defined as follows,
\bes 
&& \com{\alpha_{i},\alpha_{j}}=\com{\alpha_{i},\alpha_{j}^{\dagger}}=0, \textrm{ with  } i\neq j,  \quad
 \com{\alpha_{i},\alpha_{i}^{\dagger}}_{q^{\pm\demi}}= q^{\f{\mp N_{\alpha_{i}}}{2}}, \quad
[N_{\alpha_{i}}, \alpha_{j}^\dagger]=\delta_{{ij}}\alpha_{i}^\dagger,\quad  [N_{\alpha_{i}}, \alpha_{j}]=-\delta_{{ij}}\alpha_{i}, \label{HO}
\ees
where $[A, B]_{q^n}\equiv AB-q^n BA$. Let us point out that the operator $\alpha_{i}^\dagger \alpha_{i}$  is not the number operator  $N_{\alpha_{i}}$ but rather is equal to $[N_{\alpha_{i}}]$. From   (\ref{HO}), we have also that
\bes
&& q^{N_{\alpha_{i}}/2}\alpha_{i}^\dagger= q^{1/2}\alpha_{i}^\dagger q^{N_{\alpha_{i}}/2}, \quad q^{N_ {\alpha_{i}}/2} \alpha_{i} = q^{-1/2} \alpha_{i} q^{N_ {\alpha_{i}}/2}, \quad  \alpha_{i}^\dagger\alpha_{i} = [N_{\alpha_{i}}], \quad \alpha_{i}\alpha_{i}^\dagger= [N_{\alpha_{i}}+1].
\ees
The harmonic oscillator $\alpha_i$, $\alpha_i^\dagger$, $N_{\alpha_i}$ acts on the Fock space $F_i =\{ \sum_{n_i} c_{n_i} |n_i\ra\}$ with vacuum $|0\ra$.
\beq
\alpha_{i}|0\ra=0, \quad \alpha_i|n_i\ra = \sqrt{[n_i]} |n_i-1\ra,\, \textrm{ with } n_i\geq1, \quad \textrm{ and } \alpha_i^\dagger |n_i\ra = \sqrt{[n_i+1]} |n_i+1\ra.
\eeq
%

\medskip

The generators of $\UQ$ can be realized in terms of the pair of $q$-harmonic oscillators $(a,b)$, their adjoint and their number operator \cite{macfarl, bienden}.
 \be\label{su2qHO}
J_z=\f12 (N_a-N_b), \quad J_+=a^\dagger b, \quad J_-=b^\dagger a, \quad C= [\demi(N_{a}+N_{b})] [\demi(N_{a}+N_{b})+1].
\ee
Using this representation together with (\ref{HO}), we can  recover the commutation relations \eqref{commutationSU2q}. We can also use the Fock space  $F\sim F_{{a}}\ot F_{b}=\lbrace \sum c_{n_{a}n_{b}}|n_{a}, n_{b}\ra, \; c_{n_{a}n_{b}}\in\R \rbrace$ of this pair of $q$-harmonic oscillators to generate the representations of $\UQ$   by setting 
\beq 
j=\demi(n_{a}+n_{b}), \quad m=\demi(n_{a}-n_{b}).
\eeq 
The states $\jm$ are then homogenous polynomials in the operators $\alpha_{i}, \, \alpha_{i}^{\dagger}$.
\beq
\jm=\frac{\left(a^{\dagger}\right)^{j+m}  \left( b^{\dagger}\right)^{j-m}}{\sqrt{[j+m]! [j-m]!}}|0,0\ra.
\eeq

\section{Tensor operators for $\UQ$} \label{secTO}
%
We now introduce the concept of tensor operators. The general definition of tensor operators for a general quasitriangular Hopf algebra has been given in \cite{rittenberg}. We use their formalism in the specific case of $\UQ$. These objects are the building blocks of our construction of observables for LQG defined with $\UQ$ as gauge group. We  show  in section \ref{secResults} that the use of tensor operators  allows us to build any observables associated to an intertwiner (of a quantum or a classical group) in a straightforward manner.

\subsection{Definition and Wigner-Eckart theorem}

%
%

\begin{definition}   Tensor operators \cite{rittenberg}.\\
Let $V$ and $W$ be two representations of $\UQ$,  not necessarily irreducible, and $L(W)$ the set of linear maps on $W$.  A tensor operator $\bt$ is defined as the intertwinning linear map
\beq\label{defTO}
\begin{array}{rcl} \bt : V&\dr& L(W) \\
x&\dr& \bt(x)
\end{array}
\eeq
If we take $V\equiv V^{j}$ the irreducible representation of rank $j$ spanned by vectors $|j,m\ra$, then we note $\bt(|j,m\ra)\equiv \bt^{j}_{m}$. $\bt^j=\left(\bt^j_m \right)_{m=-j..j}$ is called a tensor operator of rank $j$. 
\end{definition}

\medskip 

A tensor operator being an intertwining map for the action of $\UQ$ means that $\bt^{j}_{m}$ transforms at the same time as an operator under the adjoint action of $\UQ$ and as a vector $\jm$. This is encoded in the {\it equivariance property}\footnote{As always we can perform the limit $q\dr 1$ to recover the tensor operators for $\su(2)$. In this case we have 
$$
[J_{z},\tjm ] = m \, \tjm, \quad   [J_{\pm},\tjm ]  = \sqrt{( j \mp m)( j\pm m +1)} \; \bt^{j}_{m\pm1}.
$$
This transformation is the infinitesimal version of 
$
g\, \tjm \, g\mone =  \sum_{m'}\rho^j_{mm'}(g)\bt^j_{m'}, \; g\in\SU(2),
$
where $\rho$ is a representation  of $\SU(2)$.}
\bes \label{TO constraints}
J_{z}\act \tjm&=& [J_{z},\tjm ] = m\, \tjm  \\
J_{\pm}\act \tjm&=& J_\pm \; \tjm \; q^{-\f{J_{z}}{2}} - q^{\pm\demi} q^{-\f{J_{z}}{2}} \; \tjm \; J_\pm = \sqrt{[ j \mp m][j\pm m +1]} \; \bt^{j}_{m\pm1}.\nn
\ees

This equivariance property has a very important consequence regarding the matrix elements of $\tjm$. 

\begin{theorem}  Wigner-Eckart theorem \cite{rittenberg}:   \label{WE}\\ 
The matrix elements   $\langle j_{1},m_{1}| \bt^{j}_{m} | j_{2},m_{2}\rangle$  are proportional to the CG coefficients.  The constant of proportionality  $N^{j}_{j_{1}j_{2}}  $ is a function of $j_{1}, \,j_{2}$ and $j$ only.
\beq
\langle j_{1},m_{1}| \bt^{j}_{m} | j_{2},m_{2}\rangle = N^{j}_{j_{1}j_{2}} \CG{c@{}c@{}c} {j& j_2& j_{1} \\ m& m_2 & m_{1}}.
\eeq
\end{theorem} 
The proof of the theorem follows from the constraints  \eqref{TO constraints} written for the matrix elements of the tensor operator. These constraints essentially implement the recurrence relations which define the CG coefficients, as given in \eqref{CGrecursion}. 

In order to have at least a  non-zero matrix element, the $j$'s in the CG coefficients must satisfy the triangular condition. This means in particular that the tensor operator does not have to be realized as a square matrix. Let us consider the cases $j=0,\demi,1$. 

\begin{itemize}
\item The scalar operator $\bt^0$ has matrix elements given in terms of $\CG{c@{}c@{}c} {0& j_2& j_{1} \\ 0& m_2 & m_{1}}$. As a consequence, we must have $j_1=j_2$ and the scalar operator must be encoded in a square matrix $(2j_1+1)\times (2j_1+1)$. 

\item The spinor operator $\bt^\demi$ matrix elements are given in terms of $\CG{c@{}c@{}c} {\demi& j_2& j_{1} \\ m& m_2 & m_{1}}$. We must have $j_2+\demi=j_1$ or $j_2-\demi=j_1$. The spinor operator cannot be realized by a square matrix. It has to be represented in terms of a rectangular matrix of either of  the type $(2j_2+2)\times (2j_2+1)$,  $(2j_2)\times (2j_2+1)$ or a direct sum of the two. 

\item In a similar way, the vector operator $\bt^1$ has matrix elements given by $\CG{c@{}c@{}c} {1& j_2& j_{1} \\ m& m_2 & m_{1}}$. Hence it must be realized as a matrix of the either of the types 
$(2j_2-1)\times (2j_2+1)$,  $(2j_2+1)\times (2j_2+1)$,  $(2j_2+3)\times (2j_2+1)$ or a direct sum of some/all of them. 
\end{itemize}

\medskip 


\subsection{Product  of tensor operators: scalar product, vector product and triple product}
We would like now to consider  the analogue of \eqref{TOrecoupling01} and \eqref{TOrecoupling0}  in terms of tensor operators. 
\begin{lemma}\label{productTO} Product of tensor operators \cite{rittenberg}.  \\ 
Let $\bt: V\dr L(W)$ and $\btt:V'\dr L(W)$ be two tensor operators then 
\beq
\begin{array}{rcl}
\bt\btt: V\ot V' &\dr& L(W) \\
(x,y)&\dr & \bt(x) \btt(y)
\end{array}
\eeq
is still a tensor operator. 
\end{lemma}
For example, we can decompose a given tensor operator in terms of two other tensor operators, using the CG coefficients. 
\be \label{TOrecoupling}
{\bt}^{ j}_m=\sum_{m_1, m_2} \,  \CG{c@{}c@{}c}{j_1& j_2&j \\ m_1& m_2 & m}   {\bf t}^{j_1}_{m_1}{\bf t}^{ j_2}_{m_2}.
\ee 
Two specific combinations will be especially relevant for us: ``scalar product'' and ``vector product''. \medskip

\subsubsection{Scalar product}\label{scalar product}
We call ``scalar product'' of two tensor operators, the projection of these operators on the trivial representation. Indeed, considering two tensor operators $\bt^{j_{1}}$ and $\btt^{j_{2}}$, we can combine them using the CG coefficients to build a tensor operator of rank 0, \ie {}a scalar operator.
\be \label{invariantTO0}
\bt^{j_{1}} \cdot \btt^{j_{2}}\equiv \sqrt{[2j_1+1]} \sum_{m_1+m_2=0}\, \CG{c@{}c@{}c}{j_1& j_2 & 0\\m_1 & m_2 & 0}\, {\bf t}^{j_1}_{m_1}  \,{\btt}^{j_2}_{m_2}=\delta_{j_1, j_2} \sum_m (-1)^{j_1-m}q^{\f{m}{2}} {\bf t}^{j_1}_{m}\,  {\btt}^{j_1}_{-m},
\ee 
In this sense, we can interpret  these quantum Clebsch-Gordan coefficients  as encoding a (non-degenerated) bilinear form $\cB^{(j)}$ defining a scalar product. 
\beq
 \cB^{(j)}(v,w)= g^{(j)}_{mn}v^mw^n=v\cdot w, \quad  g^{(j)}_{mn}=  {\sqrt{[2j_1+1]}} \CG{c@{}c@{}c}{j& j & 0\\m& n & 0} = \, \delta_{m, -n} \, {(-1)^{j-m}q^{\f{m}{2}}}\neq  g^{(j)}_{nm} .
\eeq
To have a scalar product out from a bilinear form $\cB$, we  usually demand that the bilinear form is symmetric $\cB(v,w)=\cB (\psi(v,w))$, where $\psi$ is the permutation. However due to the non-cocommutativity of the coproduct,  we have a non trivial tensor product structure.  Thus we have to discuss the symmetry with respect to  the deformed permutation $\psi_\cR= \psi \circ \cR$.  We have then 
\beq
v\cdot w= \cB(v,w) = (-1)^{2j}q^{-j(j+1)} \cB(\psi_\cR(v,w))= (-1)^{2j}q^{-j(j+1)}w\cdot v.
\eeq
We notice therefore that, modulo the factor $q^{-j(j+1)}$, if $j$ is integer we have a (deformed) symmetric bilinear form, whereas in the half integer case, it is (deformed) antisymmetric. This is consistent with the construction when $q\dr1$. Unlike in the classical case there is an extra factor $q^{-j(j+1)}$ that comes into play. 
%
%
Since we have defined a bilinear form, we can introduce the contravariant and covariant notions. If $|u\ra= \sum_m u_m|jm\ra$ is a vector (covariant object), then 
$\la u| \equiv \sum_m u_{-m}{(-1)^{j-m}q^{\f{m}{2}}} \la jm|$ will be the covector (contravariant object). This notion can be naturally extended to tensor operators. We have defined earlier the covariant tensor operators since they transform as vectors. We can introduce the contravariant tensor operators as
\beq \label{contravariantdef}
{\bf t}_j^m\equiv  {(-1)^{j-m}q^{\f{m}{2}}} \left({\bf t}^j_{-m}\right)^\dagger,
\eeq
where $\dagger$ is here the standard combination of transpose and complex conjugation.  This contravariant notion of tensor operators was actually proposed by Quesne \cite{quesne}. 

Finally, given a bilinear form, we can construct the associated notion of adjoint $\dagger_\cB$ of an operator $A$, from $\cB(A^{\dagger_\cB}v,w))=\cB(v,Aw)$. We recall that\footnote{We omit the $j$ upper index for simplicity.} $ g_{mn}=\delta_{m, -n} \, {(-1)^{j-m}q^{\f{m}{2}}}$ is antidiagonal and not symmetric, so that we need to be careful. We note $g^{mn}=(-1)^{-j-m}q^{\f{m}{2}}\delta_{-m,n}$ its inverse. Following the adjoint definition, given a bilinear form  $ g_{mn}$, we have, for a given operator $A$,  
\beq\label{transpose}
 {(A^{\dagger_\cB})^m}_n = g^{m a} {A^d}_a   g_{dn} = \left((-1)^{m-n} q^{-\f{m-n}{2}}\right) \,\,  {A_{-n}}^{-m}.
\eeq

\subsubsection{Vector product}
The notion of ``vector product'' is defined by associating a vector operator ${\hat\bt}^{1}$ to two vector operators $\bt^{1}, \,  \btt^{1}$ using   the CG coefficients, 
\beq\label{vectorproduct}
{\hat\bt}^{1}_m =(\bt^1\wedge \btt^1)_m \equiv\sum_{m_1,m_2} \CG{c@{}c@{}c}{1& 1 & 1\\ m_1 & m_2 & m}\bt^{1}_{m_1} \,  \btt^{1}_{m_2}.
\eeq
Using their value (recalled in the appendix \ref{CG app}), we obtain explicitly
\bes 
{\hat\bt}_{1} &=& \sqrt{\f{[2]}{[4]}} \left( q^{1/2} \bt^{1}_{1}\btt^{1}_{0}-q^{-1/2} \bt^{1}_{0}\btt^{1}_{1}\right), \quad 
{\hat\bt}_{-1} = \sqrt{\f{[2]}{[4]}}\left( q^{1/2} \bt^{1}_{0}\btt^{1}_{-1}-q^{-1/2} \bt^{1}_{-1}\btt^{1}_{0} \right),\nn\\
{\hat\bt}_{0} &=& \sqrt{\f{[2]}{[4]}} \left(  \bt^{1}_{1}\btt^{1}_{-1}- \bt^{1}_{-1}\btt^{1}_{1} + \left(q^{\demi} - q^{-\demi}\right)\bt^{1}_{0}\btt^{1}_{0} \right).\nn
\ees
As we shall see when giving a realization of the vector operators, this  vector product is related to the commutation relations of the $\su(2)$ algebra (when $q=1$) and  to  Witten's proposal describing the $q$-deformation of the $\su(2)$ algebra \cite{witten}. 
Combining the scalar product with the wedge product, we obtain the generalization of the triple product. 
\beq
(\bt^{1}\wedge \bt^{1} )\cdot  \bt^{1}\equiv \sum_{m_i}  (-1)^{1-m_3}q^{\f{m_3}{2}} \CG{c@{}c@{}c}{1& 1 & 1\\ m_1 & m_2 & m_3} \bt^{1}_{m_1} \,  \bt^{1}_{m_2}  \bt^{1}_{-m_3}. 
\eeq
This is nothing else than the image of trivalent intertwiner $\eqref{trivalent}$ when restricted to $j_1=j_2=j_3=1$.  The generalization to any $j_i$ is then 
\beq
(\bt^{j_1}\wedge \bt^{j_2} )\cdot  \bt^{j_3} \equiv  
\sum_{m_i}  (-1)^{j_3-m_3}q^{\f{m_3}{2}}\CG{c@{}c@{}c}{j_1& j_2 & j_3\\ m_1 & m_2 & m_3}\bt^{j_1}_{m_1} \,  \bt^{j_2}_{m_2}  \bt^{j_3}_{-m_3}.  
\eeq
In general, given a set of tensor operators,  we can use the relevant  intertwiner coefficients, to construct a scalar operator out of them. Observables for an intertwiner will be the generalization of this construction.

\subsection{Tensor products of tensor operators}
The tensor product of tensor operators necessitates more attention. Indeed if $\bt\in L(W)$ and $\btt\in L(W')$  are tensor operators for $\UQ$, then in general $\bt\ot\btt$ will not be a tensor operator for $\UQ$. To see this, first we recall that we need the coproduct to define the action of the generators $J_{\alpha}$ on $|j_{1}m_{1},j_{2}m_{2}\ra$. For example, 
\be
\cop J_{+} |j_{1}m_{1},j_{2}m_{2}\ra = \left(J_{+}\ot K + K\mone \ot J_{+}\right)|j_{1}m_{1},j_{2}m_{2}\ra, \quad K\equiv q^{\f{J_{z}}{2}}.
\ee
If $\bt\ot\btt$ is a (linear) module homomorphism, we have then 
\bes\label{mince}
\left(\bt\ot\btt\right) (\cop J_{+} |j_{1}m_{1},j_{2}m_{2}\ra) &=&\left(\bt\ot\btt \right)(  J_{+}\ot K + K\mone \ot J_{+}|j_{1}m_{1},j_{2}m_{2}\ra) \nn\\
 &=& (J_{+}\act \tjma )\ot( K \act \btt^{j_2}_{m_2}) + (K\mone \act \tjma) \ot (J_{+}\act \btt^{j_2}_{m_2}).
\ees
On the other hand this is must be equal to the action of $J_{+}$ on $\bt\ot\btt $ seen as a linear map $V\ot V'\dr W\ot W'$, so that 
\beq\label{boudu}
J_{+}\act (\bt\ot\btt ) = {\left(J_{+}\right)_{V\ot V'}}\, (\bt\ot\btt )\, {\left(K\mone\right)_{W\ot W'}} - q^{\demi} {\left(K\mone\right)_{V\ot V'}}\, (\bt\ot\btt )\, {\left(J_{+}\right)_{W\ot W'}}. 
\eeq
We recall that by definition we have 
\bes
{\left(K^{\pm1}\right)_{W\ot W'}} &=&  \cop K ^{\pm1}= {\left(K^{\pm 1}\right)_{W}} \ot {\left(K^{\pm1}\right)_{W'}} \\
{\left(J_{+}\right)_{W\ot W'}}&=& \cop J_{+}= {\left(J_{+}\right)_{W}} \ot {\left(K\right)_{W'}} + {\left(K\mone\right)_{W}} \ot  {\left(J_{+}\right)_{W'}}.
\ees
If $\bt\ot\btt $ is a tensor operator, we must have \eqref{mince} = \eqref{boudu}, which gives (we omit for simplicity the indices) 
\beq\label{TOtensor}
 && (J_{+} \bt K\mone -q^{\demi} K\mone  \bt J_{+} )\ot( K  \btt K\mone) + (K\mone  \bt K) \ot (J_{+} \btt K\mone - q^{\demi} K\mone \btt J_{+} )= \\
 &&J_{+}\bt K\mone \ot K\btt K\mone + K\mone \bt K\mone \ot J_{+}\btt K\mone-q^{\demi} \left( K\mone \bt J_{+}\ot K\mone \btt K+ K\mone \bt K\mone \ot K\mone \btt J_{+} \right).\nn
 \ees
If $\btt=\one$, then \eqref{TOtensor} is satisfied for any $\bt$ but when $\bt=\one$ and $\btt \neq \one$, the constraint  \eqref{TOtensor} is not satisfied in general\footnote{Note that in the limit $q\dr 1$, this would be satisfied. Hence $\one\ot \btt$ is a tensor operator for $\su(2)$.}. The problem can be identified with the non-commutativity of the coproduct \cite{rittenberg}.  Indeed, the operator $\one\ot \bt$ can be seen as obtained from the permutation of $\bt\ot\one$, but since we are dealing now with a non-commutative tensor product, we need to consider the deformed permutation $\psir$ instead of $\psi$.

\begin{lemma}\label{TOR}\cite{rittenberg}
If  $\bt$  is a tensor operator of rank $j$ then $\, ^{(1)}\bt=\bt \ot \one$ and  $\, ^{(2)}\bt=\psi_{\cR}(\bt \ot \one) \psi_{\cR}\mone= \cR_{21} (\one\ot \bt)\cR_{21}\mone$  are tensor operators of rank $j$
\end{lemma}
We extend the construction to an arbitrary number of tensor products\footnote{$\cR_{ms}=  \one^{s-1}\ot \cR_2\ot \one ^{m-s-1}\ot \cR_1$, using notations of \eqref{Rmatrix}.}. 
\beq \label{gros boulot}
 ^{(i)}\bt=\cR_{ii-1} \cR_{ii-2}.. \cR_{i\,1} (\one\ot\one\ot ..\ot \one\ot \bt)\cR_{i\,1}\mone.. \cR_{ii-2}\mone \cR_{ii-1} \mone \ot \one\ot ..\ot \one.
\eeq
By abuse of notation, we say that $\, ^{(i)}\bt$ acts on the $i^{th}$ Hilbert space, even though it is not really the case when $q\neq 1$. Note also that if $q=1$, tensor operators which act on different Hilbert spaces will commute, but when $q\neq1$, this will not be the case in general due to the presence of the $\cR$-matrices. 

When we consider the scalar product of  tensor operators living acting on the same Hilbert space, the $\cR$-matrices disappear which simplifies the calculations. 
\begin{lemma}\label{lemma norm}
The  scalar product of  the tensor operators $\,^{(i)}\bt^{j_{1}}$ and $\,^{(i)}\btt^{j_{2}}$  can be reduced to 
\beq
\,^{(i)}\bt^{j_{1}}\cdot \,^{(i)}\btt^{j_{2}} = \one \ot..\one\ot \bt^{j_{1}}\cdot \btt^{j_{2}} \ot\one\ot..\ot \one. 
\eeq
\end{lemma}
This lemma simply follows from \eqref{gros boulot}.

\section{Realization of tensor operators of rank $1/2$ and $1$ for $\UQ$}\label{realization}
The abstract theory of tensor operators has been summarized above. We want to illustrate the construction by giving some realization of these tensor operators. We know that any representation $V^{j}$ of $\UQ$ can be recovered from the fundamental spinor representation $\demi$ and the CG coefficients. In the same way,  the most important operators to identify are the spinor operators. If we know them, we can concatenate them using the CG coefficients to obtain any other tensor operators. We first present the realization of the spinor operators using $q$-harmonic oscillators and then present the vector operators  realized in terms of either the $q$-harmonic oscillators or the $\UQ$ generators.

 \subsection{Rank $1/2$ tensor operators}
Rank $\demi$ tensor operators (\ie spinor operators) $\bt^{\demi}_{m}$ should be solution of the following constraints.
 \be\label{spinorop}
J_{\pm}\triangleright \mathbf{t}^{\demi}_{\mp}= \mathbf{t}^{\demi}_{\pm}, \quad J_{\pm}\triangleright \mathbf{t}^{\demi}_{\pm}=0, \quad J_z \triangleright \mathbf{t}^{\demi}_{\pm}=\pm \f12 \mathbf{t}_\pm^{\demi}.
\ee
Using the Schwinger-Jordan realization of $\UQ$ generators given in \eqref{su2qHO}, we can solve these equations and we get two solutions $T^{\demi}$  and  $\tilde{T}^{\demi}$ satisfying \eqref{spinorop}.
 \be \label{T12op}
T^{\demi}=\left(\tabl{c}{A^\dagger\\ B^\dagger} \right)= \left(\tabl{c}{a^\dagger q^{N_a/4}\\   b^\dagger q^{(2N_a+N_b)/4} }\right), \qquad \tilde{T}^{\demi}=\left(\tabl{c}{\tilde{B}\\ \tilde{A}} \right) = \left(\tabl{c}{q^{(2N_a+N_b+1)/4}b\\ -q^{(N_a-1)/4}a }\right).
\ee
We recall that $a$ and $b$ are $q$-harmonic oscillators which satisfy the modified commutation relations  \eqref{HO}. 
We can check that  $T^{\demi}$ and $\tilde{T}^{\demi}$ are Hermitian conjugate to each other, according to the  modified bilinear form we have defined in Section \ref{scalar product} (see \eqref{contravariantdef}). 
When looking at the limit $q\dr 1$, we have  
\be \label{T12opq=1}
T^{\demi} \, \rightarrow \, \tau^{\demi}=\left(\tabl{c}{a^\dagger\\ b^\dagger} \right), \qquad \tilde{T}^{\demi} \rightarrow \, \tilde{\tau}^{\demi}=\left(\tabl{c}{b\\ -a} \right).
\ee

\medskip

This explicit realization of the tensor operators allows to check  explicitly  the Wigner-Eckart theorem, and to identify the normalization of the operators through this realization. In particular, for the $q$-deformed spinor operator matrix elements, we have 
\bes\label{spinor normalization}
\la j_{1}, m_{1} | T^{\demi}_m | j_2, m_2 \ra &=&    \delta_{j_{1}, j_2+1/2}\, N^\demi_{j_2}\, _qC^{1/2 \, j_2 \, j_{1}}_{m \; m_2 \,m_{1}}, \textrm{ with } N^\demi_{j_2}= \left( ([d_{j_2}])^{1/2} q^{\f{j_2}{2}}\right) , \nn\\ 
\la j_{1}, m_{1} | \tilde{T}^{\demi}_m | j_2, m_2 \ra &=& \delta_{j_{1}, j_2-1/2} \,\tilde N^\demi_{j_2} \, _qC^{1/2 \, j_2 \, j_{1}}_{m \; m_2 \,m_{1}},   \textrm{ with } \tilde N^\demi_{j_2}= \left(  ([d_{j_2}])^{1/2} q^{\f{1}{4}(2j_2-1)}\right), 
\ees
where $m=\pm 1/2$ and $d_{j}=2j+1$. We have therefore the two possible realizations of spinor operators in terms of rectangular matrices. Note that the above choice of normalization $N^\demi_{j_2}$ and $ \tilde N^\demi_{j_2}$ can be modified because the spinor operators  $T^\demi$ and $\tilde T^\demi$ are defined up to a multiplicative function of $N_a+N_b$.  Therefore, $N^\demi_{j_2}$ and $ \tilde N^\demi_{j_2}$ can be any function of $ j_2$.


\medskip
To form observables for a $N$-valent intertwiner, we need to define  spinor operators built from the tensor product of N spinor operators. The explicit realization of the tensor product of spinor operators has been discussed in details by Quesne \cite{quesne}. The calculation amounts to calculate  \eqref{gros boulot} for an arbitrary number $N$ of tensor products, in the case of the spinor operators $\bt^{\demi}$.

We outline now the outcome of this calculation and give the expression of these spinor operators in terms of the $q$-deformed harmonic oscillators $a_i^\dagger, \, a_i, \, N_{a_i}, \, b_i^\dagger, \, b_i, \, N_{b_i} \, \in F_i \sim F_{a_i} \otimes F_{b_i}$ where the $F_i \, (i=1, \cdots, N)$ are $N$ independent $q$-Fock spaces. Let us define the tensor operators $^{(i)}T^{\demi}$ and $^{(i)}\tilde{T}^{\demi}$ living in $\mathcal{F}\equiv (\otimes_{i=1}^N F_{a_i})(\otimes_{i=1}^N F_{b_i})$ which ``act'' on the $i^{th}$ Hilbert space. 
\be \label{T12i}
^{(i)}T^{\demi}=\left( \tabl{c}{\cA^\dagger_i\\ \cB_i^\dagger}\right), \qquad \,^{(i)}\tilde{T}^{\demi}=\left( \tabl{c}{\tilde{\cB}_i \\ \tilde{\cA}_i} \right), \qquad \textrm{ for } i \in \{ 1, \cdots, N\}, 
\ee
where 
\be \tabl{l}{
\,^{(i)}T^{\f12}_+:=\cA^\dagger_i=(\displaystyle{\otimes_{k=1}^{i-1}} q^{\f{N_{a_k}-N_{b_k}}{4}})\, a_i^\dagger q^{\f{N_{a_i}}{4}}, \\
\\
\,^{(i)}T^{\f12}_-:=\cB^\dagger_i=(\displaystyle{\otimes_{k=1}^{i-1}} q^{\f{-N_{a_k}+N_{b_k}}{4}})\, b_i^\dagger q^{\f{2N_{a_i}+N_{b_i}}{4}} + (q^{\f14}-q^{-\f34}) \left[\displaystyle{\sum_{l=1}^{i-1}} (\displaystyle{\otimes_{k=1}^{l-1}} q^{\f{-N_{a_k}+N_{b_k}}{4}})\, a_l b_l^\dagger (\displaystyle{\otimes_{k=l+1}^{i-1}} q^{\f{N_{a_k}-N_{b_k}}{4}})\right] \, a_i^\dagger q^{\f{N_{a_i}}{4}}, \\
\\
 \,^{(i)}\tilde{T}^{\f12}_+:=\tilde{\cB}_i= (\displaystyle{\otimes_{k=1}^{i-1}} q^{\f{N_{a_k}-N_{b_k}}{4}})\,q^{\f{2N_{a_i}+N_{b_i}+1}{4}}\,b_i, \\
 \\
  \,^{(i)}\tilde{T}^{\f12}_-:= \tilde{\cA}_i= (\displaystyle{\otimes_{k=1}^{i-1}} q^{\f{-N_{a_k}+N_{b_k}}{4}})\, (-q^{\f{N_{a_i}-1}{4}}a_i )+ (q^{\f14}-q^{-\f34}) \left[\displaystyle{\sum_{l=1}^{i-1}} (\displaystyle{\otimes_{k=1}^{l-1}} q^{\f{-N_{a_k}+N_{b_k}}{4}})\, a_l b_l^\dagger (\displaystyle{\otimes_{k=l+1}^{i-1}} q^{\f{N_{a_k}-N_{b_k}}{4}})\right] \,q^{\f{2N_{a_i}+N_{b_i}+1}{4}}b_i.
}\nn \ee 
These operators will be the building blocks of our construction of $\UQ$-observables presented in the following section. It will be necessary to have their explicit form in terms of the harmonic oscillators in order to recover the $\UUQn$ structure in Section \ref{uuqn}. 

\medskip

Note that if $i\neq 1$, the two spinor operators  $^{(i)}T^{\demi}$ and $^{(i)}\tilde{T}^{\demi}$ are \textit{no more} Hermitian conjugated to each other.  Indeed, 
$
 (\cA_i^\dagger)^\dagger \neq-q^{1/4}\tilde{\cA}_i, \; (\cB_i^\dagger)^\dagger \neq q^{-1/4}\tilde{\cB}_i, \; i \in \{ 2, \cdots, N\}.
$
 To emphasize this lack of Hermiticity, we introduce the notation,
\be
\cC_i\equiv -q^{1/4} \tilde{\cA}_i, \qquad \cD_i\equiv q^{-1/4}\tilde{\cB}_i, \qquad  \forall \, i \, \in \{1,\cdots, N\}.
\ee
That is, we can rewrite the spinor operators $^{(i)}\tilde{T}^{\demi}$ as $^{(i)}\tilde{T}^{\demi}= \left( \tabl{c}{q^{\f14}\cD_i \\ -q^{-\f14}\cC_i} \right)$. Quesne has calculated all possible commutation relations between the components of $^{(i)}T^{\demi}_{\pm }$, $^{(j)}\tilde T^{\demi}_{\pm}$ for any $i,\, j \, \in \{1, \cdots, N\}$ \cite{quesne}. 
 First let us give the commutation relations when $1\leq i= j \leq N$.
\bes \label{qcommutations1}
&& \cB_i^\dagger \cA_i^\dagger=q^{1/2}\cA^\dagger_i\cB^\dagger_i,   \quad \cC_i\cD_i=q^{1/2}\cD_i\cC_i, \quad   \cD_i\cA_i^\dagger=q^{1/2} \cA^\dagger_i \cD_i,\quad  \cC_i\cB^\dagger_i=q^{1/2}\cB^\dagger_i \cC_i,   \quad \cC_i\cA^\dagger_i= q  \cA^\dagger_i \cC_i +1, \nn \\
 &&  \cD_i\cB^\dagger_i=q\cB^\dagger_i \cD_i+ (q-1) \cA^\dagger_i\cC_i+1. 
\ees 
When $1\leq i<j \leq N$, we have 
\bes\label{qcommutations2}
&&\cA_i^\dagger \cA_j^\dagger= q^{-1/4} \cA^\dagger_j \cA^\dagger_i, \quad \;  \cA^\dagger_i \cB^\dagger_j=q^{1/4} \cB_j^\dagger \cA_i^\dagger- (q^{3/4}-q^{-1/4})\cA^\dagger_j \cB^\dagger_i, \quad   \cB^\dagger_i \cA_j^\dagger=q^{1/4} \cA_j^\dagger \cB^\dagger_i, \quad \; \cB_i^\dagger \cB_j^\dagger= q^{-1/4} \cB^\dagger_j \cB^\dagger_i, \nn \\
&& \nn \\
&&\cD_i \cD_j= q^{-1/4} \cD_j \cD_i, \quad  \cD_i\cC_j=q^{1/4} \cC_j \cD_i- (q^{3/4}-q^{-1/4})\cD_j\cC_i, \quad  \cC_i \cD_j=q^{1/4} \cD_j \cC_i, \quad  \cC_i \cC_j= q^{-1/4}\cC_j \cC_i, \nn\\
&& \nn \\
&&\cA^\dagger_i \cD_j= q^{-1/4}\cD_j \cA^\dagger_j, \quad \; \cD_i\cA^\dagger_j= q^{-1/4} \cA^\dagger_j \cD_i, \quad \cB^\dagger_i \cC_j= q^{-1/4} \cC_j \cB^\dagger_i, \quad \; \cC_i \cB^\dagger_j=q^{-1/4}\cB^\dagger_j \cC_i,   \quad   \cB_i^\dagger \cD_j=q^{1/4}\cD_j \cB^\dagger_i, \nn   \\
&& \nn\\
&& \cD_i\cB_j^\dagger=q^{1/4}\left(\cB_j^\dagger \cD_i +(1-q^{-1})\cA^\dagger_j \cC_i \right),  \quad\cA^\dagger_i \cC_j=q^{1/4} \left(\cC_j \cA_i^\dagger + (q-1)\cD_j\cB^\dagger_i \right),   \quad  \cC_i \cA_j^\dagger= q^{1/4} \cA^\dagger_j \cC_i. 
\ees
These commutation relations are quite cumbersome and they illustrate  that the components of operators acting on different Hilbert spaces do not commute when $q\neq1$. Obviously, when $q=1$, they simplify a lot. 

\subsection{Rank $1$ tensor operators}\label{vector operator}
Rank $1$ tensor operators (\ie vector operator) for $\UQ$ have been identified \cite{rittenberg}. These operators are important as in the LQG context, they will encode the notion of flux operator. We explicitly construct them and provide their commutation relations, when they act on different legs, or  not.

We can construct them using the spinor operators $T^{\demi}$, $\tilde{T}^{\demi}$ and the CG coefficients.
\beq
\bt^{1}_{m}= \sum_{m_{1},m_{2}}\CG{c@{}c@{}c}{\f12 & \f12 & 1\\ m_{1} & m_{2} & m}T^{\demi}_{m_{1}}\tilde{T}^{\demi}_{m_{2}}.
\eeq
Using the explicit non-zero CG coefficients given in the appendix \ref{CG app},  we have that
\bes
&&\bt^1_{\pm1}=T^{\f12}_{\pm} \tilde{T}^{\f12}_{\pm},\quad  \bt^1_0\,=\f{1}{\sqrt{[2]}} \left( q^{-\f14} T^{\f12}_{+} \tilde{T}^{\f12}_{-} + q^{\f14}T^{\f12}_{-} \tilde{T}^{\f12}_{+} \right).
\ees
Explicitly, 
we obtain that 
\bes
\bt^{1}_{1}&=&{q^{-\f12}} \,q^{\f{(3N_a+N_b)}{4}} a^\dagger b ={q^{-1/2}}q^{\f12 (N_a+N_b)}q^{\f{J_z}{2}}J_+, \\
\bt^{1}_0&=&- \f{1}{[2]^\demi} \left( q^{-1}q^{N_a/2} [N_a] -q^{N_a+N_b/2}[N_b]\right)= -\f{q^{-1/2}}{[2]^\demi}q^{\f12 (N_a+N_b)}(q^{-1/2}J_+J_--q^{1/2}J_-J_+), \\
\bt^1_{-1}&=&- {q^{-\demi}} \, q^{(3N_a+N_b)/4} b^\dagger a= - {q^{-\demi}}q^{\f12 (N_a+N_b)}q^{\f{J_z}{2}}J_-.
\ees
Once again, we can check that the Wigner-Eckart is satisfied,
\be\label{normalization vector}
\la j_1, m_1 | \bt^{1}_l|j_2, m_2\ra= \delta_{j_1,j_2}\,  N^1_{j_2}   \, \CG{c@{}c@{}c}{1& j_2& j_1 \\ l& m_2 &  m_1} \quad \textrm{ with } N^1_{j_2}=- q^{j_2-\f{1}{2}} \left(\f{\left([2j_2] [2j_2+2]\right)}{[2]}\right)^\demi,
 \ee
and $l \in \, \{ -1, 0,1 \}.$ In this realization, the vector operator is realized as a square matrix. Note that the normalization $N^1_{j_2}$  comes here from the  chosen spinor  normalization \eqref{spinor normalization}. For a given vector operator, we can always consider an arbitrary normalization $N^1_{j_2}$. 

\medskip

An important remark is that in the limit $q\dr1$, the components of the vector operator become proportional to the components of the $\su(2)$-generators, 
\beq\label{vector op su2}
\bt^{1} \, \rightarrow \tau^1=\left( \begin{array}{c} J_+\\-\sqrt{2} J_z\\ -J_-\end{array}\right).
\eeq
That is the $\su(2)$ generators are very simply related to vector operators. Let us now go back to our definition of generalized scalar product \eqref{invariantTO0} and generalized vector product \eqref{vectorproduct}. In the $q=1$-case, the $q$-deformed CG coefficients of equations \eqref{invariantTO0} and \eqref{vectorproduct} are simply replaced by the standard $\su(2)$ CG coefficients. In particular the scalar product is still the projection on the trivial rank and we can define the ``norm" of the vector operator $\tau^1$, given by $\tau^1 \cdot \tau^1\equiv \sum_{m_1+m_2=0}\,  C^{\,1\; \; \,1 \;\; \;0}_{m_1 \, m_2 \, 0}\, \tau^1_{m_1} \, \tau_{m_2}^1 $. This simplifies into
\be \label{su2Norm}
\tau^1 \cdot \tau^1=-2 \vec{J} \cdot \vec{J}
\ee
where the $\su(2)$ set of generators $\vec J$ is seen as a 3-vector with components $J_x=\f12 (J_++J_-), \, J_y=\f{1}{2i} (J_+-J_-)$ and $J_z=J_z$ and the $\cdot$ of the left-hand side of \eqref{su2Norm} denotes the standard scalar product of 3-vectors. That is, in the non-deformed case, the norm of the vector operator is proportional to the quadratic Casimir of $\su(2)$, $\cC=\vec{J} \cdot \vec{J}$. 
The norm of the $\UQ$ vector operator is by definition a $\UQ$-invariant but it is not proportional to $|\vec{J}|^2$ anymore. Indeed,
\be
t^1\cdot t^1 \propto (q^{\f12}-q^{-\f12})^2 J_-^2J_+^2 + ([2J_z+4]-[2J_z])J_-J_++[2J_z+2][2J_z]
\ee
where the proportionality coefficient is a function of $q^{\f{N_a+N_b}{2}}$.

The ``vector product" operation in the case $q=1$ can be understood  as the commutator of the $\su(2)$ generators, which is also the natural way to encode the notion of vector product, as used in LQG.  Indeed
\bes\label{vector prod su2}
(\tau^1 \wedge \tau^1)_1 &=& \f{1}{\sqrt{2}} \left(  \tau^{1}_{1}\tau^{1}_{0}- \tau^{1}_{0}\tau^{1}_{1}\right)= [J_z,J_+]= J_+ = {\tau}_{1}^1 ,\nn\\
(\tau^1 \wedge \tau^1)_{-1}& =& \f{1}{\sqrt{2}}\left(  \tau^{1}_{0}\tau^{1}_{-1}- \tau^{1}_{-1}\tau^{1}_{0} \right) = [J_z,J_-] =-J_-= {\tau}_{-1}^1 ,\nn\\
(\tau^1 \wedge \tau^1)_0&=& \f{1}{\sqrt{2}} \left(  \tau^{1}_{1}\tau^{1}_{-1}- \tau^{1}_{-1}\tau^{1}_{1} \right)= \f{1}{\sqrt{2}} [J_-,J_+] = -\sqrt{2}J_z= {\tau}_{0} ^1.
\ees
We see therefore that this vector product can be understood as the commutator of the $\su(2)$ generators, which is also the natural way to encode the notion of vector product, as used in LQG. 

One can check explicitly using the above realization of the vector operators when $q\neq1$ that 
\bes \label{vector prod su2q}
({\bt}^1\wedge {\bt}^1)_{1} &=& \sqrt{\f{[2]}{[4]}} \left( q^{1/2} \bt^{1}_{1}\bt^{1}_{0}-q^{-1/2} \bt^{1}_{0}\bt^{1}_{1}\right), \quad 
({\bt}^1\wedge {\bt}^1)_{-1} = \sqrt{\f{[2]}{[4]}}\left( q^{1/2} \bt^{1}_{0}\bt^{1}_{-1}-q^{-1/2} \bt^{1}_{-1}\bt^{1}_{0} \right),\nn\\
({\bt}^1\wedge {\bt}^1)_{0} &=& \sqrt{\f{[2]}{[4]}} \left(  \bt^{1}_{1}\bt^{1}_{-1}- \bt^{1}_{-1}\bt^{1}_{1} + \left(q^{\demi} - q^{-\demi}\right)\bt^{1}_{0}\bt^{1}_{0} \right).
\ees
Thus, in the quantum group case, the vector product of vector operators is different than the commutation relations defining $\UQ$. The matrix elements of this new vector operator can be expressed in terms of the matrix elements of ${\bt}^1$,
\be
\la j, \, m_1 | ({\bt}^1\wedge {\bt}^1)_{\alpha } | j, \, m_2\ra = \f{[2j-1]-[2j+3]}{[2]([4])^{\f12}} q^{j-\f12}\la j, \, m_1 | {\bt}^1_{\alpha } | j, \, m_2\ra
\ee

We see therefore that in the classical case when $q=1$, the generators are related to vector operators and different structures, such as the adjoint action, the commutator and vector product are encoded in the same way. When $q\neq1$, all these different degeneracies are actually lifted. Let us summarize in the table below all the possible relations in the cases of $\su(2)$ and $\UQ$.


\begin{center}
\begin{tabular}{|c|c|c|}
\hline \T & $\su(2)$ & $\UQ$ \B \\
\hline \T Generators & $J_\sigma$, $\sigma=\pm,z$ & $J_\sigma$, $\sigma=\pm,z$ \\ & with commutation relations & with commutation relations \\
&$[J_+, J_-]=2J_z$ & {\bf $[J_+, J_-]=\f{q^{J_z}-q^{-J_z}}{q^{\f12}-q^{-\f12}}$}  \\
&$[J_\pm, J_z]=\mp J_\pm$ & $[J_\pm, J_z]=\mp J_\pm$\B \\
\hline \T Vector operators & $\tau^1=\left( \begin{array}{c} J_+\\-\sqrt{2} J_z\\ -J_-\end{array}\right)$ & {\bf $\bt^1\propto \left(\begin{array}{c}q^{\f{J_z}{2}}J_+ \\ -\f{1}{\sqrt{[2]}}(q^{-1/2}J_+J_--q^{1/2}J_-J_+) \\ -q^{\f{J_z}{2}}J_- \end{array} \right)$ }\B \\
\hline \T Adjoint action & $J_\sigma \act \cO = [J_\sigma, \cO]$ for $\sigma= \pm, \, z$ & $ J_\pm \act \cO = J_\pm \cO q^{-\f{J_z}{2} } -q^{\pm \f12} q^{-\f{J_z}{2}} \cO J_\pm, \; J_z \act \cO= [J_z, \cO]$\B \\ 
\hline \T ``Scalar product" & $\tau^1 \cdot \tau^1=-2\cC=-2 |\vec{J}|^2$   & ${\bt}^1 \cdot {\bt}^1= I$ where $I$ is a $\UQ$-invariant;  \\
($\cdot$ defined by \eqref{invariantTO0})& where $\cC$ is the quadratic Casimir of $\su(2)$. &  $|\vec{J}|^2$ is not a Casimir for $\UQ$.  \B\\
\hline \T ``Vector product" &  $(\tau^1\wedge \tau^1)_{\pm 1}=[J_z, J_\pm],$ & $({\bt}^1\wedge {\bt}^1)_\alpha = \hat{{\bt}}^1_\alpha=$ vector operator; not simply related to \\
($\wedge$ defined by \eqref{vectorproduct}) & $ (\tau^1\wedge \tau^1)_{z}=\f{1}{\sqrt{2}} [J_-, J_+].$&  the commutators between generators of $\UQ$.  \B \\
\hline
\end{tabular}
\end{center}


\medskip

The extension of $\bt^1$ to $\,^{(i)}\bt^1$, for $i \in \{1, \cdots, N\}$, can be done either through \eqref{gros boulot} or by using the spinor operators $^{(i)}T^{\demi}_{m_{1}}$ and $ \,^{(i)}\tilde{T}^{\demi}_{m_{2}}$ given in \eqref{T12i}.
\beq
^{(i)}\bt^{1}_{m}= \sum_{m_{1},m_{2}}\CG{c@{}c@{}c}{\f12 & \f12 & 1\\ m_{1} & m_{2} & m}\,^{(i)}T^{\demi}_{m_{1}} \,^{(i)}\tilde{T}^{\demi}_{m_{2}} \dr \left\{ \begin{array}{l} ^{(i)}\bt^1_1= q^{\f14}\cA^\dagger _i \cD_i=\cA^\dagger_i \tilde{\cB}_i \\ ^{(i)}\bt^1_0=\f{1}{\sqrt{[2]}} \left(  q^\demi \cB^\dagger_i \cD_i - q^{-\demi}\cA^\dagger_i \cD_i\right)=\f{1}{\sqrt{[2]}} \left(  q^{\f14} \cB^\dagger_i \tilde{\cB}_i + q^{-\f14} \cA^\dagger_i \tilde{\cA}_i \right) \\^{(i)}\bt^1_{-1}= -q^{-\f14}\cB_i ^\dagger \cC_i = \cB^\dagger_i \tilde{\cA}_i \end{array}\right.
\eeq
Explicitly, in terms of the $\UQ$-generators, we have,
\bes
\,^{(i)}{\bt}^1_{1}&=&q^{\sum_{k=1}^{i-1} \,^{(k)}{J_z }} \,^{(i)}{J_+} q^{\f{\,^{(i)}J_z}{2}} q^{\f{N_{a_i}+N_{b_i}}{2}}, \nn \\
\,^{(i)}{\bt}^1_{0}&=&\f{1}{\sqrt{[2]}} [ -q^{-\f12} (q^{-\f12} \,^{(i)}J_+\,^{(i)}J_- -q^{\f12} \,^{(i)}J_- \,^{(i)}J_+) q^{\f{N_{a_i}+N_{b_i}}{2}} \nn\\
&& \qquad \qquad + (q^{\f12}-q^{-\f12})(1+q^{-\f12}) \sum_{l=1}^{i-1} [q^{(\f{\,^{(l)}J_z}{2}+\sum_{k=l+1}^{i-1}\,^{(k)}J_z)}\,^{(l)}J_-]\,^{(i)}J_+ q^{\f{\,^{(i)}J_z}{2}} q^{\f{N_{a_i}+N_{b_i}}{2}}] \nn \\
\,^{(i)}{\bt}^1_{-1}&=&-q^{-1} q^{-\sum_{k=1}^{i-1}(\,^{(k)}J_z) }\,^{(i)}J_- q^{\f{^{(i)}J_z}{2}} q^{\f{N_{a_i}+N_{b_i}}{2}} \nn\\
&& \qquad \qquad -q^{-1}(q^{\f12}-q^{-\f12}) \sum_{l=1}^{i-1} \left( q^{-\sum_{k=1}^{l-1}(\,^{(k)}J_z+\f{\,^{(l)}J_z}{2})}\,^{(l)}J_-\right) (q^{-\f12} \,^{(i)}J_+\,^{(i)}J_- -q^{\f12} \,^{(i)}J_- \,^{(i)}J_+)q^{\f{N_{a_i}+N_{b_i}}{2}} \nn \\
&& \qquad \qquad+q^{-1}(q^{\f12}-q^{-\f12})^2 \left(\sum_{l=1}^{i-1} q^{-\sum_{k=1}^{l-1}\f{^{(k)}J_z}{2}} \,^{(l)}J_- q^{\sum_{k=l+1}^{i-1}\f{^{(k)}J_z}{2}} \right)^2 \,^{(i)}J_+ q^{\f{\,^{(i)}J_z}{2}} q^{\f{N_{a_i}+N_{b_i}}{2}} \nn
\ees

With this choice of normalization inherited from \eqref{normalization vector}, the commutation relations between the $^{(i)}\bt^{1}_{m}$ are quite complicated. For $1\leq i < j \leq N$,
\bes \label{commutationT1}
&&\,^{(i)}{\bt}^1_{1} \,^{(j)}{\bt}^1_{1}=q^{-1} \,^{(j)}{\bt}^1_{1} \,^{(i)}{\bt}^1_{1} \qquad \qquad  \quad \qquad \,^{(i)}{\bt}^1_{1} \,^{(j)}{\bt}^1_{0}=\,^{(j)}{\bt}^1_{0} \,^{(i)}{\bt}^1_{1} + (q^{-1}-q) \,^{(j)}{\bt}^1_{1} \,^{(i)}{\bt}^1_{0} \nn \\
&&    \,^{(i)}{\bt}^1_{1} \,^{(j)}{\bt}^1_{-1}= q \,^{(j)}{\bt}^1_{-1} \,^{(i)}{\bt}^1_{1}-(q-1)[2]\,^{(j)}{\bt}^1_{0} \,^{(i)}{\bt}^1_{0}+(q^2-1)(1-q^{-1})\,^{(j)}{\bt}^1_{1} \,^{(i)}{\bt}^1_{-1} \nn \\
&&\,^{(i)}{\bt}^1_{0} \,^{(j)}{\bt}^1_{1}= \,^{(j)}{\bt}^1_{1} \,^{(i)}{\bt}^1_{0} \qquad \qquad \qquad \quad \quad \; \,^{(i)}{\bt}^1_{0} \,^{(j)}{\bt}^1_{0}= \,^{(j)}{\bt}^1_{0} \,^{(i)}{\bt}^1_{0} - (q^{\f12}-q^{-\f12})(q+1)(1+q^{-1})\,^{(j)}{\bt}^1_{1} \,^{(i)}{\bt}^1_{-1} \\
&& \,^{(i)}{\bt}^1_{0} \,^{(j)}{\bt}^1_{-1}= \,^{(j)}{\bt}^1_{-1} \,^{(i)}{\bt}^1_{0}+(q^{-1}-q)\,^{(j)}{\bt}^1_{0} \,^{(i)}{\bt}^1_{-1} \nn \\
&& \,^{(i)}{\bt}^1_{-1} \,^{(j)}{\bt}^1_{1}= q \,^{(j)}{\bt}^1_{1} \,^{(i)}{\bt}^1_{-1} \qquad \qquad \quad \quad \quad \;  \,^{(i)}{\bt}^1_{-1} \,^{(j)}{\bt}^1_{0}=\,^{(j)}{\bt}^1_{0} \,^{(i)}{\bt}^1_{-1} \quad \qquad \qquad \,^{(i)}{\bt}^1_{-1} \,^{(j)}{\bt}^1_{-1}=q^{-1}\,^{(j)}{\bt}^1_{-1} \,^{(i)}{\bt}^1_{-1} \nn
\ees
For $i=j \in \{1, \cdots, N\}$,
\bes
&& \,^{(i)}{\bt}^1_1 \,^{(i)}{\bt}^1_0=q^{-1} \,^{(i)}{\bt}^1_0 \,^{(i)}{\bt}^1_1 +\f{q^{\f12}-q^{-\f12}}{\sqrt{[2]}} \cE_{ii} \,^{(i)}{\bt}_1^1 +q^{-1} \sqrt{[2]} \,^{(i)}{\bt}_1^1, \nn \\
&& \,^{(i)}{\bt}^1_1 \,^{(i)}{\bt}^1_{-1}= q^{-1} \,^{(i)}{\bt}^1_{-1} \,^{(i)}{\bt}^1_{1} + \f{(q^{-1}-1)}{[2]} (\,^{(i)}{\bt}^1_0)^2 +\f{2(q^{\f12}-q^{-\f12})}{[2]^{\f32}} \,^{(i)}{\bt}_0^1 \cE_{ii}+\f{2q^{-1}}{\sqrt{[2]}} \,^{(i)}{\bt}_0^1 - \f{q^{-1}(q^{\f12}-q^{-\f12})}{[2]}\cE_{ii} \\
&& \,^{(i)}{\bt}^1_0 \,^{(i)}{\bt}^1_{-1}=q^{-1} \,^{(i)}{\bt}^1_{-1} \,^{(i)}{\bt}^1_{0} +\f{q^{\f12}-q^{-\f12}}{[2]} \cE_{ii} +q^{-1}\sqrt{[2]} \,^{(i)}{\bt}_{-1}^1,
\ees
where $\cE_{ii}:=-q^{\f14} \cA_i^\dagger \tilde{\cA}_i +q^{-\f14} \cB^\dagger_i \tilde{\cB}_i$ is a $\UQ$-invariant (see section \ref{uuqn}) and it commutes with any $\,^{(i)}{\bt}^1_\alpha, \; (\alpha=\pm, z)$.

\section{Observables for the intertwiner space} \label{secResults}
As emphasized in the introduction, we focus on the quantum group $\UQ$ with  $q$ real, which is  relevant for 3d Euclidian gravity with $\Lambda<0$ and the physical case, \ie 4d Lorentzian gravity with $\Lambda>0$. 

\subsection{General construction and properties of intertwiner observables}\label{tricks}
From now on, we consider the space of $N$-valent intertwiners with $N$ legs ordered from 1 to $N$. 
Let's consider $n$ tensor operators $\,^{(\alpha)}\bt^{J_\alpha}$ of respective rank $J_\alpha$, associated with the $\alpha^{th}$ leg of the vertex, built from \eqref{gros boulot}. 
To construct an observable, \ie a scalar operator, we can use the same combination that would appear in the definition of an intertwiner built out from the vectors $|J_\alpha,m_\alpha\ra$. 
 Indeed, if 
$
|\iota_{J_{1}..J_{n}}\ra= \sum_m c^{J_1..J_n}_{m_1..m_n}|J_1m_1,..J_nm_n\ra, 
$
then,
\beq
I^{J_{1}..J_{n}}= \sum_{m_i} c^{J_1..J_n}_{m_1..m_n}\,^{(1)}\bt^{J_1}_{m_1}..\,^{(n)}\bt^{J_n}_{m_n}
\eeq
will be a scalar operator. 
Like for intertwiners, the bivalent and trivalent ones are the simplest and we can write them explicitly, 
\bes
&&  I^{J_\alpha J_\beta}\equiv \,^{(\alpha)}\bt^{J_\alpha}\cdot \,^{(\beta)}\bt^{J_\beta}=   \delta_{J_\alpha ,J_\beta} \sum_m (-1)^{J_\alpha-m}q^{\f{m}{2}} \,^{(\alpha)}{\bf t}^{J_\alpha}_{m}\,   \,^{(\beta)}{\bt}^{J_\beta}_{-m}\equiv I^{J_\alpha }_{\alpha\beta},\,
\\
&& I^{J_\alpha J_\beta J_\gamma}\equiv(\,^{(\alpha)}\bt^{J_\alpha}  \wedge  \,^{(\beta)}\bt^{J_\beta})\cdot   \,^{(\gamma)}\bt^{J_\gamma} =  \sum_{m_i} {(-1)^{J_\gamma-m_3}q^{\f{m_3}2}} \CG{c@{}c@{}c}{J_{\alpha}& J_{\beta} & J_{\gamma}\\m_{1}& m_{2}& m_{3}}\,^{(\alpha)}\bt^{J_\alpha}_{m_1}  \,^{(\beta)}\bt^{J_\beta}_{m_2} \,^{(\gamma)}\bt^{J_\gamma}_{-m_3}.
\ees
We recognize  the  generalized notions of  respectively the scalar product and the triple product. 

\medskip

This construction works well for operators acting on an intertwiner, however in the general LQG context, we need to deal with spin networks, so we need to consider the tensor product of such intertwiners $|\iota_{j_{1}..j_{N}}\ra\ot |\iota'_{j_{1'}..j_{N'}}\ra\ot...$.  Although the tensor product is not commutative, we do not need to use the deformed permutation to define an operator acting on any intertwiner of the tensor product. Indeed, since an intertwiner is a $\UQ$-invariant vector, the tensor product involving such invariant vectors is commutative.  

More explicitly, we have seen earlier that if $\bt$ is a tensor operator, then $\one\ot \bt$ will not be in general a tensor operator. However if  $\one\ot \bt$  is restricted to act on some invariant vectors $|\iota\ra \ot | \iota'\ra$, then $\one\ot \bt$ will still be a tensor operator. 

To see this, let us consider the invariant vectors $|\iota\ra\ot | \iota'\ra \in W\ot W'$. We recall that an invariant vector means that 
\beq\label{inv vector}
J_\pm |\iota\ra= 0, \quad K^{\pm 1}|\iota\ra = |\iota\ra. 
\eeq
Let us first determine the transformation of $\one\ot \bt$ as a representation of $\UQ$, that is  \eqref{mince}, when acting on the vectors $|\iota\ra\ot | \iota'\ra$  
\beq\label{123}
&&\left((J_{+}  K\mone -q^{\demi} K\mone  J_{+} )\ot K  \btt K\mone + \one  \ot (J_{+} \btt K\mone - q^{\demi} K\mone \btt J_{+}) \right)|\iota\ra \ot |\iota'\ra = \one\ot (J_+ \act \btt)|\iota\ra \ot |\iota'\ra 
\eeq
If $\one\ot \bt$ transforms well when restricted to the invariant vectors $|\iota\ra\ot | \iota'\ra$, we must recover the same outcome as \eqref{123} when considering $\one\ot \bt$ transforming as an operator, that is  \eqref{boudu}.
\beq
&& \left(J_{+} K\mone \ot K\btt K\mone + \one \ot J_{+}\btt K\mone-q^{\demi} \left( K\mone  J_{+}\ot K\mone \btt K+ K^{-2} \ot K\mone \btt J_{+} \right)\right)|\iota\ra \ot |\iota'\ra\nn\\ 
&&=
\one\ot (J_+ \act \btt)|\iota\ra \ot |\iota'\ra.
\eeq
In the right hand side of the two above equations, we have used \eqref{inv vector} to recover $\one\ot (J_+ \act \btt)|\iota\ra \ot |\iota'\ra$. A similar calculation can be made for the action of $J_-$ and $K$. Hence, the operator $\one\ot \btt$ transforms as a tensor operator of same rank as $\btt$ when restricted to act on an invariant state $|\iota\ra \ot |\iota'\ra$. This means that we can just focus on the observables associated to one intertwiner, and if we look at another intertwiner a priori we do not need to order the vertices, unless we look at observables that live on both intertwiners at the same time.

\medskip

If we have many legs in our intertwiner, it might cumbersome to calculate the terms  $\,^{(i)}\bt^J$ and $\,^{(j)}\bt^J$   and then calculate the observable $I^J_{ij}$, since we have to use extensively the deformed permutations and a lot of CG coefficients (or $\cR$ matrices) appear then. If we know the matrix elements of $I^J_{12}$ and $I^{J}_{21}$, we can construct by induction all the other terms. We know that by definition
\bes
I^J_{12}&=& \sqrt{[2J+1]}\sum_{m_i} \CG{c@{}c@{}c}{J&J&0\\m_1&m_2&0} \, (\bt^{J}_{m_1} \otimes \one ) \cR_{21}(\one\ot  \bt^{J}_{m_2} )\cR\mone_{21},\\
I^J_{13}&=&\sqrt{[2J+1]}\sum_{m_i} \CG{c@{}c@{}c}{J&J&0\\m_1&m_2&0} \, (\bt^{J}_{m_1} \otimes \one\ot\one )\cR_{32} \cR_{31}(\one\ot \one\ot \bt^{J}_{m_2} )\cR\mone_{31}\cR\mone_{32},\\
I^J_{23}&=& \sqrt{[2J+1]}\sum_{m_i} \CG{c@{}c@{}c}{J&J&0\\m_1&m_2&0}\, \cR_{21} (\one\ot \bt^{J}_{m_1} \otimes \one  ) \cR_{21}\mone \, \cR_{32} \cR_{31}(\one\ot \bt^{J}_{m_2} )\cR\mone_{31}\cR\mone_{32}.
\eeq
We can construct the observable $I^{J}_{13}$ from $I^J_{12}$,  by permuting $2$ with $3$, using   the braided permutation $\psi_{23}$ defined in \eqref{deformedPerm}. Upon this permutation, we have in particular that $\cR_{21}$ becomes $\cR_{31}$.
\bes
 \psi_{23} I^J_{12} \psi_{23}\mone &=& \sqrt{[2J+1]}\sum_{m_i}  \cR_{32} \CG{c@{}c@{}c}{J&J&0\\m_1&m_2&0}\, (\bt^{J}_{m_1} \otimes \one \ot \one ) \cR_{31}(\one\ot \one\ot \bt^{J}_{m_2} )\cR\mone_{31} \cR_{32}\mone \nn\\
&=& \sqrt{[2J+1]}\sum_{m_i} \CG{c@{}c@{}c}{J&J&0\\m_1&m_2&0}\, (\bt^{J}_{m_1} \otimes \one \ot\one ) \cR_{32} \cR_{31}(\one\ot \one\ot \bt^{J}_{m_2} )\cR\mone_{31} \cR_{32}\mone\nn\\
&=&  I^J_{13}. 
\ees
We have used the fact that $\cR_{23}$ and $\bt^J_{m_1} \otimes \one \ot\one$ commute. This can be extended  to arbitrary $I^J_{1j}$. Now we would like to consider the construction of $I^J_{23}$ from $I^J_{12}$. As a matter of fact, we can start from $I^J_{13}$ and permute 1 and 2 using the deformed permutation $\psi_{12}$. 
\bes
 \psi_{12} I^J_{13}  \psi_{12}\mone=  \sqrt{[2J+1]}\sum_{m_i} \CG{c@{}c@{}c}{J&J&0\\m_1&m_2&0}\, \cR_{21}( \one \otimes \bt^{J}_{m_1}  \ot \one ) \cR_{31} \cR_{32}(\one\ot \one\ot \bt^{J}_{m_2} )\cR\mone_{32} \cR_{31}\mone \cR_{21}\mone.
\ees
To simplify this expression, we use the Yang-Baxter equation,  
\beq\label{yangbaxter}
\cR_{dc}\cR_{db}\cR_{cb}= \cR_{cb}\cR_{db}\cR_{dc},
\eeq
with $c=2, \, b=1, d=3$.
We have then 
\bes
 \psi_{12} I^J_{13}  \psi_{12}\mone & =&  \sqrt{[2J+1]}\sum_{m_i} \CG{c@{}c@{}c}{J&J&0\\m_1&m_2&0}\, \cR_{21} ( \one \otimes \bt^{J}_{m_1}  \ot \one ) \cR_{31} \cR_{32}(\one\ot \one\ot \bt^{J}_{m_2} )\cR\mone_{21} \cR_{31}\mone \cR_{32}\mone. \nn\\
&=&  \sqrt{[2J+1]}\sum_{m_i} \CG{c@{}c@{}c}{J&J&0\\m_1&m_2&0}\, \cR_{21} ( \one \otimes\bt^{J}_{m_1}  \ot \one ) \cR_{31} \cR_{32} \cR\mone_{21} (\one\ot \one\ot \bt^{J}_{m_2} ) \cR_{31}\mone \cR_{32}\mone
\ees
where we used that $\cR_{21}\mone$ commutes with $\one\ot \one\ot \bt^{J}_{m_2}$. We use again the Yang-Baxter equation \eqref{YB1}, for the product of $\cR$ matrices in the middle of the above expression. 
\beq\label{YB1}
 \cR_{21}\mone \cR_{32}\cR_{31}\cR_{21}=\cR_{31}\cR_{32}.
\eeq
\bes
 \psi_{12} I^J_{13}  \psi_{12}\mone &=& \sqrt{[2J+1]}\sum_{m_i} \CG{c@{}c@{}c}{J&J&0\\m_1&m_2&0} \, \cR_{21} ( \one \otimes \bt^{J}_{m_1}  \ot \one )   \cR_{21}\mone \cR_{32}\cR_{31}\cR_{21} v\mone_{21} (\one\ot \one\ot \bt^{J}_{m_2} ) \cR_{31}\mone \cR_{32}\mone \nn\\
&=& \sqrt{[2J+1]}\sum_{m_i} \CG{c@{}c@{}c}{J&J&0\\m_1&m_2&0} \, \cR_{21} ( \one \otimes \bt^{J}_{m_1}  \ot \one )   \cR_{21}\mone \cR_{32}\cR_{31} (\one\ot \one\ot \bt^{J}_{m_2} ) \cR_{31}\mone \cR_{32}\mone \nn\\
&=&  I^J_{23}.
\ees
This is the relevant expression for $I^J_{23}$. Hence, we can obtain any $I^J_{ij}$ with $i<j$, using the braided permutation, starting from $I^J_{12}$. A similar argument applies to construct the terms $I^J_{ji}$ with $i<j$. We can obtain them by induction on the braided permutation starting from the first term $I^J_{21}$.

\medskip

Now that we have provided a general rule and some tricks to construct observables, it is natural to answer the following questions.
\begin{itemize}
\item Can we generate any observables from a fundamental algebra of observables?
\item What is the physical meaning and implications of some of the key-observables defined in the $\UQ$ context?
\end{itemize} 
We explore these questions now.

\subsection{$\UUQn$ formalism for LQG defined over $\UQ$}\label{uuqn}
We want to construct the "smallest" observables. It is therefore natural to consider the observables built from the scalar product of spinor operators \eqref{T12i}. Since we have two types of spinor operators, we have different possible combinations. 
\bes
\cE_{\alpha\beta}\!\equiv\! - \, ^{(\alpha)}T^{\demi} \cdot ^{(\beta )}\tilde{T} ^{\demi}, \quad 
 \cG ^{\dagger}_{\alpha\beta}\!\equiv\! \/  - ^{(\alpha)}T ^{\demi} \cdot ^{(\beta )}T^{\demi}  \quad 
\cF_{\alpha\beta} \!\equiv\!  \/  - ^{(\alpha)}\tilde{T}^{\demi} \cdot ^{(\beta )}\tilde{T}^{\demi}. 
\ees
Note that since the operators on different legs do not commute, we could a priori choose a different order of $T$ and $\tilde{T}$ in the definition of $\cE_{\alpha\beta}$. One can show however that choosing the order leads to the same operator modulo a constant factor. This factor comes from the (deformed) symmetry  of the scalar product as well as the commutation relations between the spinor operators acting on different legs.

Let us focus  on the operators $\cE_{\alpha\beta}$. Consider first the spinor operators which  act on the same leg $\alpha=\beta=i$. 
\be
\cE_{ii}:=- \,^{(i)}T^{\demi} \cdot \,^{(i)}\tilde{T}^{\demi}=-q^{\f14} \cA_i^\dagger \tilde{\cA}_i +q^{-\f14} \cB^\dagger_i \tilde{\cB}_i.
\ee
Having in mind Lemma \ref{lemma norm}, we can forget about the tensor product, and the only relevant action is on the leg $i$ so 
\beq
\cE_{ii} |\iota_{j_1...j_N}\ra= [2j_i]|\iota_{j_1...j_N}\ra.
\eeq
Consider now the spinor operators which  act on different legs $i$ and $j$.  
\beq
\cE_{ij}= \cA_i^\dagger {\cC}_j + \cB^\dagger_i {\cD}_j.
\eeq
The action of $\cE_{12}=\cA_1^\dagger {\cC}_2 + \cB^\dagger_1 {\cD}_2$ on a trivalent intertwiner is given by 
\bes
\cE_{12}|\iota_{j_1j_2j_3}\ra&=& -\tilde{N}^{\f12}_{j_2} N^{\f12}_{j_1} (-1)^{j_1+j_2+j_3-1} q^{-\f34 j_1} \sqrt{[2j_1+2][2j_2]} \left\{\tabl{ccc}{j_2-\f12 & \f12 & j_2 \\
j_1 & j_3 & j_1 + \f12} \right\} |i_{j_1+\f12 j_2-\f12 j_3}\ra 
\ees
with the normalization choice
\be
N^{\f12}_j=[d_j]^{\f12} q^{\f{j}{2}}, \qquad \tilde{N}^{\f12}_j=[d_j]^{\f12} q^{\f{j}{2}-\f14}.
\nn \ee
The other operators $\cE_{ij}$ $(i, \, j \in \{1, 2, 3\}$) can be constructed using the tricks described in the previous section.
In a similar way, we get 
\bes
{q^{\f34}\cF_{12}}|i_{j_1 j_2 j_3}\ra&=&- \tilde{N}^{\f12}_{j_2} \tilde{N}^{\f12}_{j_1} (-1)^{j_1+j_2+j_3} q^{\f12( j_2+1)} \sqrt{[2j_1][2j_2]} \left\{\tabl{ccc}{j_2-\f12 & \f12 & j_2 \\
j_1 & j_3 & j_1 - \f12} \right\} |i_{j_1-\f12 j_2-\f12 j_3}\ra \nn \\
{q^{\f14}\cG^\dagger_{12}} |i_{j_1 j_2 j_3}\ra&=& -N^{\f12}_{j_2} N^{\f12}_{j_1} (-1)^{j_1+j_2+j_3} q^{-\f12( j_1+\f34)} \sqrt{[2j_1+2][2j_2+2]} \left\{\tabl{ccc}{j_2+\f12 & \f12 & j_2 \\
j_1 & j_3 & j_1 + \f12} \right\} |i_{j_1+\f12 j_2+\f12 j_3}\ra \nn \\
\ees

\medskip

When we perform the limit $q\dr 1$, the operators $\cA_i^\dagger$, $\cB_i^\dagger$, $\cC_i$, $\cD_i$ become respectively $a_i^\dagger$, $b_i^\dagger$, $a_i$ and $b_i$, that is  the standard harmonic oscillators operators.
Hence in this limit, the operators $\cE_{ij}$ become $a^{\dagger}_{i} a_{j} +  b^{\dagger}_{i} b_{j} $  which are the generators $E_{ij}$   of a $\u(n)$ Lie algebra, written using the Schwinger-Jordan representation. In a similar way, the operators $\cF_{ij}, \, \cG_{ij}$ become respectively $F_{ij}$ and $ F^{\dagger}_{ij} $ defined as follows
\bes
 \cG ^{\dagger}_{ij}  \stackrel{q\dr 1}{\dr}    a^{\dagger}_{i} b^{\dagger}_{j} -  b^{\dagger}_{i} a^{\dagger}_{j}  \!=\! F^{\dagger}_{ij} \label{Fdag}, \quad 
\cF_{ij} \stackrel{q\dr 1}{\dr}   a_{i} b_{j} -  b_{i} a_{j} \!=\!  F_{ij}. \label{F}
\ees 
We recognize the operators $E, F, F^\dagger$ which are the basis of the $\U(N)$ formalism \cite{un0,un1,un2}.  They appear very naturally in our framework. 

\medskip 

It is then natural to demand if the operators $\cE_{ij}$ are the generators $\UUQn$.
First, let us recall  the definition of $\UUQn$ Cartan Weyl generators  \cite{BiedenharnBook}.  We have respectively  the raising, diagonal, lowering  operators $\fE_{ii+1}$, $\fE_{i}$, $\fE_{i-1i}$, with the following commutation relations
\beq\nn
\com{\fE_{ii}, \fE_{jj}}=0, \quad \com{\fE_{ii},\fE_{jj+1}}=(\delta_{ij}-\delta_{ij+1})\fE_{jj+1}, \quad \com{\fE_{ii},\fE_{j-1j}}=(\delta_{ij+1}-\delta_{ij})\fE_{j-1j}, \quad \com{\fE_{ii+1},\fE_{j-1j}}=\delta_{ij}(\fE_{i}-\fE_{i+1}).
\eeq
The other generators are constructed by induction.
\bes
\fE_{ij}&=&q^{\demi \fE_{j-1}}\left(\fE_{ij-1}\fE_{j-1j}-q^\demi \fE_{j-1,j}\fE_{ij-1}  \right), \quad j>i+1 \\
\fE_{ji}&=&q^{-\demi \fE_{j-1}}\left(\fE_{jj-1}\fE_{j-1i}-q^{-\demi} \fE_{j-1,i}\fE_{jj-1}  \right), \quad j>i+1. 
\ees
Note that $\fE_{ij}$ is not necessarily the adjoint of $\fE_{ji}$ due to the presence of $q$. The coproduct is defined 
as follows
\bes \label{}
\cop \fE_i= \fE_i\ot \one + \one \ot \fE_i, \quad \cop \fE_{ii+1}=\fE_{ii+1} \ot q^{\fE_i+\fE_{i+1}}+ q^{\fE_i+\fE_{i+1}}\ot \fE_{ii+1}, \quad .
\ees
The coproduct for the other generators are obtained by induction. 

\medskip

The Schwinger-Jordan map allows to express these generators in terms of $N$ $q$-harmonic oscillators $a_i$. 
\beq
\fE_{ij}=a_ia_j^\dagger, \quad \fE_i=\demi (N_i - N_{i+1})
\eeq
To have the representation of these generators in terms of $N$ pairs of $q$-harmonic oscillators $(a_i,b_i)$, we use the coproduct:
\bes\label{2ho}
&& \fE_{ii}:=N_{a_i} + N_{b_i}, \\ \nn
&& \fE_{i, i+p}:= a^\dagger_i a_{i+p}\, q^{\f{N_{b_i}+2(\sum_{l=1}^{p-1}N_{b_{i+l}})-N_{b_{i+p}}}{4}}  +q^{\f{-N_{a_i}+2(\sum_{l=1}^{p-1}N_{a_{i+l}})+N_{a_{i+p}}}{4}} b^\dagger_i b_{i+p} \\
&& \qquad \qquad \qquad+ (q^{-\f14}-q^{\f34}) \sum_{k=1}^{p-1}\left( q^{\f{N_{a_{i+k}}+2(\sum_{l=k+1}^{p-1}N_{a_{i+l}})+N_{a_{i+p}}}{4}}a^\dagger_i a_{i+k} \, q^{\f{N_{b_i}+2(\sum_{l=1}^{k-1}N_{b_{i+l}})+N_{b_{i+k}}}{4}}b_{i+k}^\dagger b_{i+p}\right),\\
&& \fE_{i+p,i}:=a_i a_{i+p}^\dagger \, q^{\f{N_{b_i}-2(\sum_{l=1}^{p-1}N_{b_{i+l}})-N_{b_{i+p}}}{4}}  +q^{\f{-N_{a_i}-2(\sum_{l=1}^{p-1}N_{a_{i+l}})+N_{a_{i+p}}}{4}} b_i b_{i+p}^\dagger \nn \\
&& \qquad \qquad \qquad+ (q^{\f14}-q^{-\f34}) \sum_{k=1}^{p-1}\left( q^{\f{-N_{a_{i}}-2(\sum_{l=1}^{k-1}N_{a_{i+l}})-N_{a_{i+k}}}{4}}a^\dagger_{i+p} a_{i+k} \, q^{\f{-N_{b_{i+k}}-2(\sum_{l=k+1}^{p-1}N_{b_{i+l}})-N_{b_{i+p}}}{4}}b_{i+k}^\dagger b_{i}\right).
\ees 
Using the definition of the $\fE_{ij}$ in terms of the $q$-harmonic oscillators $(a_i,b_i)$ deduced from the expression of the spinor operators, we   can identify  a non linear relationship between these $\fE_{ij}$ and the  $\cE_{ij}$.
\bes\label{relation un}
&&\cE_{ii}=q^{-\f12} q^{\f{\fE_{ii}}{2}} [\fE_{ii}], \quad  \cE_{i,i+1}=q^{\f{\fE_{i+1,i+1}}{4}} \fE_{i,i+1}, \quad  \cE_{i+1,i}=q^{\f{\fE_{ii}}{2}}\fE_{i+1,i} \,q^{\f{\fE_{i+1,i+1}}{4}} , \quad   \cE_{i,i+p}= q^{\f{-\sum_{l=1}^{p-1} \fE_{i+l, i+l}+\fE_{i+p,i+p}}{4}} \fE_{i, i+p},\nn \\ 
&& \quad  \cE_{i+p,i}= q^{\f{2\fE_{ii}+\sum_{l=1}^{p-1} \fE_{i+l, i+l}}{4}} \fE_{i+p, i}\, q^{\f{\fE_{i+p,i+p}}{4}}. 
\ees
To have a non-linear redefinition of the generators is something common when dealing with quantum groups. For example, there exists different realizations of $\UQ$, all related with a non-linear redefinition of the generators \cite{Curtright:1989sw}. Biedenharn recalls also different definitions of the generators of $\UUQn$ related by nonlinear transformations in \cite{BiedenharnBook}. For some choice of generators, the commutation relation might take a simpler shape but the coproduct would be more complicated, and vice-versa. The key point is here that we have  found that the intertwiner carries a representation of $\UUQn$, and this generalizes the results of \cite{un0,un2}. 

\medskip

 In the classical case, when $q=1$, it was shown that the intertwiner carries an \textit{irreducible} $\u(n)$ representation \cite{un1}. We can wonder whether a similar result also holds here. The answer is positive. A  cumbersome proof can probably be obtained by looking at the Casimirs of $\UUQn$. We do not want to follow this route. Instead we would like to recall the seminal  results  by Jimbo, Rosso and Lustzig \cite{Jimbo, Rosso, Lusztig} which essentially state that all the finite dimensional representations of the deformation $\cU_q(\g)$ of the enveloping algebra $\cU(\g)$, where $\g$ is any complex simple Lie algebra, are completely reducible. The irreducible representations can be classified in terms of highest weights and in particular they are  \textit{deformations} of the irreducible representations of   $\cU(\g)$, when $q$ is \textit{not} root of unity. We can extend  this result to the semi-simple case and to $\UUQn$ in particular (see Section 2.5 of \cite{BiedenharnBook} for example). Now we know that when $q=1$, the intertwiner is an irreducible representation of $\u(n)$, hence by deforming the enveloping algebra, the representation of $\UUQn$ carried by the $\UQ$ intertwiner must stay an irreducible  representation. As a consequence, the $\UQ$ intertwiner must carry an irreducible representation of $\UUQn$, just as in the classical case.

\medskip

Finally, we can discuss the hermiticity property of the scalar operators we have constructed. Indeed, we expect an observable to be self-adjoint. The operators $\cE_{ij}$ are not self-adjoint, but this should not come as a surprise. Indeed the classical operators $E_{ij}$ are not hermitian either. However, the adjoint $\left(E_{ij}\right)^\dagger = E_{ji}$ is still a generator. This means that we can do a linear change of basis $E_{ij}\dr E_{ij}+ \left(E_{ij}\right)^\dagger$ in the $\u(n)$ basis to construct self-adjoint generators. This is actually how the formalism was initially introduced in \cite{un0}. The Cartan Weyl generators $\fE_{ij}$  when expressed in terms of the harmonic oscillators satisfy a similar property, namely $\fE_{ij}^\dagger=\fE_{ji}$ \cite{BiedenharnBook}. As a consequence, from the $\cE_{ij}$, we can do a (non-linear) change of basis and construct the relevant hermitian $\UUQn$ generators which will be $\UQ$ invariant,  using the maps \eqref{relation un}.

 \section{Geometric interpretation of  some observables in  the LQG context}\label{lqg-result}
In LQG with $\Lambda=0$, the intertwiner is understood as the fundamental chunk of quantum space. For a 2d space, it is dual to a face, whereas in 3d it is dual to a polyhedron. The intertwiner is invariant under the action of $\su(2)$, hence the observables should be invariant under the  adjoint action of $\su(2)$. We see that the use of tensor operators allows to construct in a direct manner such observables: we need to construct operators which transform as a scalar under the adjoint action of $\su(2)$. 
We have seen in the previous section how this formalism can be extended to the quantum group case $\UQ$  in a direct manner. When $\Lambda=0$, some observables have a clear geometrical meaning. We have for example the quantum version of the angle, the length... 
We now explore  the generalization of these geometric operators in 3 dimensions, in the Euclidian case  with $\Lambda<0$.  

\medskip
 
For simplicity we are going to focus on the three-leg intertwiner. When $\Lambda=0$,  we know that it encodes the quantum state of a triangle. Let us recall quickly the main geometric features of a triangle, either flat or hyperbolic.

Classically a \textit{flat} triangle can be described by the normals $\vec n_i$, $i=a,b,c$ to its edges, such that $|\vec n_i|=\ell_i$ is the edge length. To have a triangle,  the normals need to sum up to zero, this is the closure constraint.  All the geometric information of the triangle can then be expressed in terms of these normals, as recalled in the table below. 

\begin{center}
\begin{figure}[h]
\includegraphics[scale=.8]{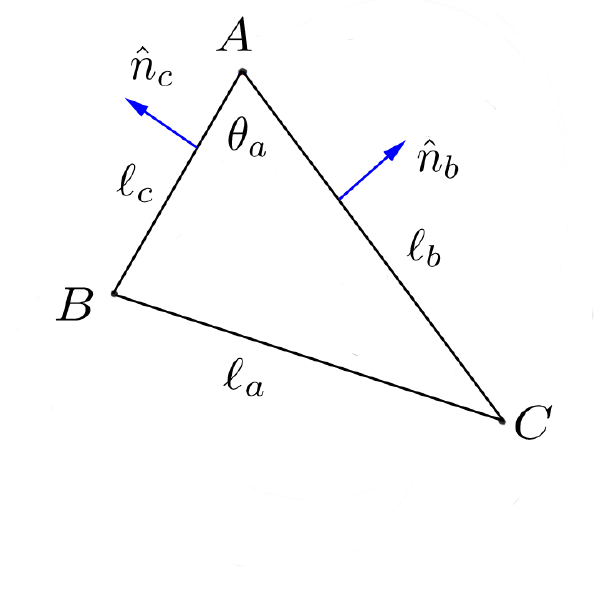}
\quad \includegraphics[scale=.8]{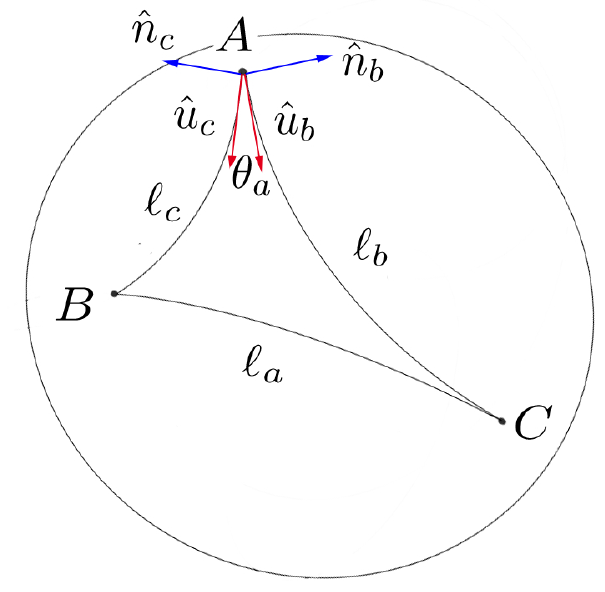}
\caption{The hyperbolic triangle is represented in the Poincar\'e disc. The (outgoing) normals $\hat n_i$ are defined in the tangent plane at the vertex of the triangle, as the orthogonal vectors to the tangent vectors $\hat u_i$.}
\end{figure}
\end{center}
Let us consider now an hyperbolic triangle.  Its edges are geodesics in the 2d hyperboloid of radius $R$. Unlike the flat triangle, an hyperbolic triangle can be characterized by its three angles $\theta_i$ or the three lengths $\ell_i$ of its edges. The hyperbolic cosine laws relate the edge lengths and the angles (see the table below). The area $\cA$ of the triangle is given in terms of the angles. 
\beq\label{hyper area}
\cA=(\pi-(\theta_a+\theta_b+\theta_c))R^2.
\eeq
In order to make easier the limit to the flat case, we can encode all this information in terms of the normals.  Note  however that due to the curvature,  we have a different tangent space at each point of the edge. The tangent vectors and their normal  are therefore not living in the same vector space for different points.   In the curved case, we shall consider the normals $\vec n_i$ at each vertex of the triangle. As a direct consequence, the closure constraint in the curved case is  subtler than in the flat case. We postpone the study of this constraint to a detailed  analysis of the relevant phase space in \cite{phase space}. We recall  in the following table the main geometric features of the flat and hyperbolic triangles, in terms of the normals. We use the notation $s=\demi(\ell_a+\ell_b+\ell_c)$.
\smallskip
\begin{center} 
\begin{tabular}{|c|c|c|}
\hline &Flat case, $\Lambda=0$& Hyperbolic case, $\Lambda<0$, $R=|\Lambda|^{-\demi}$ 
\\\hline
\T \textrm{Closure constraint: } & $\sum_i \vec n_i =0$ & To be determined \cite{phase space} 
\B \\ \hline
\T \B\textrm{Edge length:  } &$|\vec n_i|=\ell_i$ &$|\vec n_i|=\sinh \f{\ell_i}R$  
\\\hline
\T \B \textrm{Cosine law: } &$\cos\theta_a =-\hat n_b \cdot \hat n_c = -\frac{\ell_a^2  - \ell_b^2-\ell_c^2}{2\ell_b\ell_c} $&  $\cos \theta_{a} = -\hat n_b \cdot \hat n_c 
= \frac{-\cosh \frac{\ell_a}{R} + \cosh \frac{\ell_b}{R} \cosh \frac{\ell_c}{R}}{\sinh \frac{\ell_b}{R}\sinh \frac{\ell_c}{R}}$
\\ \hline
\T \B \textrm{Area:  } &  $\cA^2=\f14\left( s(s-\ell_a)(s-\ell_b)(s-\ell_c)\right)$&  \hspace{2mm} $\sin^2 \f{\cA}{2R^2}= \f{\sinh(\f{s}{2R})\sinh(\f{s-\ell_a}{2R})\sinh(\f{s-\ell_b}{2R})\sinh(\f{s-\ell_c}{2R})}{\cosh^2\f{\ell_a}{2R}\cosh^2\f{\ell_b}{2R}\cosh^2\f{\ell_c}{2R}}$\hspace{2mm} \\ \hline
\end{tabular} 
\smallskip

\end{center}
The quantization of the flat triangle can be done very naturally. The quantum state is given by the three-leg $\SU(2)$ intertwiner. We associate to normalized normals  $\vec n_i$ to the flux operators $\,^{(i)}\vec J$, which we know now to be related to the $\SU(2)$ vector operators $\,^{(i)}\tau^1$ (cf  section \ref{vector operator}).  This provides a direct quantization of  all the geometric data: closure constraint, length, angles, area (see \cite{valentin-laurent} for a recent review of these results).

\medskip

We consider now a $\UQ$ three-leg intertwiner $| \iota_{j_bj_cj_a}\ra$. The ordering we choose for the legs is fixed as we have already emphasized before. We would like to check whether it encodes the quantum state of an hyperbolic triangle. We use the $\UQ$ tensor operators to probe the geometry of this state of geometry. Since we are in the 3d framework with a negative cosmological constant, we take $q=e^{\lambda}$, with $\lambda=\f{\ell_p}{R}$, and $\Lambda=-R^2$.

\medskip

\paragraph*{Angle operator.}
Since we know that the angles specify completely  the hyperbolic triangle, we can focus first on operators characterizing angles. By analogy with the non-deformed case, we define the scalar product of the vector operators $\,^{(i)}\hat \bt^1$ and $\,^{(j)}\hat \bt^1$, with chosen normalization $\hat N^1_{j_i}=1$ and $i\neq j$.  We look at the  action of this operator on the three-leg intertwiner $| \iota_{j_bj_cj_a}\ra$. For simplicity we focus on $ \,^{(b)}\hat \bt^1 \cdot  \,^{(c)}\hat \bt^1$, since we know how to recover the other types of operators from this one using tricks developed in Section \ref{tricks}.  
\bes\label{hyper}
 \,^{(b)}\hat \bt^1 \cdot  \,^{(c)}\hat \bt^1 \, |\iota_{j_{b}j_{c}j_{a}}\ra 
 &=& - q \,  \frac{\cosh\frac{\lambda}{2} \cosh ((j_a+\demi) \lambda) - \cosh ((j_b+\demi)\lambda) \cosh((j_c+\demi)\lambda))}{\sqrt{(\sinh (j_b\lambda ))(\sinh ((j_b+1)\lambda))(\sinh (j_c\lambda))(\sinh ((j_c+1)\lambda))}} |\iota_{j_{b}j_{c}j_{a}}\ra, \nn\\
 &=& - q \,    \frac{\cosh\frac{\lambda}{2} \cosh ((j_a+\demi) \lambda) - \cosh ((j_b+\demi)\lambda) \cosh((j_c+\demi)\lambda))}{\sqrt{(\sinh^2((j_b+\demi)\lambda) - \sinh^2\f\lambda2)(\sinh^2((j_c+\demi)\lambda) - \sinh^2\f\lambda2)}} |\iota_{j_{b}j_{c}j_{a}}\ra,  
\ees 
where we have used $q=e^\lambda$ and $ \sinh (j\lambda))(\sinh ((j+1)\lambda))=\sinh^2((j+\demi)\lambda) - \sinh^2\f\lambda2$. We recognize in \eqref{hyper} a quantization of the hyperbolic cosine law, provided we  consider the quantization of the length edge given by $\ell\dr (j+\demi)\ell_p$. Note that the factors $\sinh^2\f\lambda2$ in the denominator and $\cosh\frac{\lambda}{2}$ in the numerator can be interpreted as  ordering ambiguity factors, arising from the respective quantization of $\sinh \f{\ell_i}{R}$ and $\cosh \f{\ell_i}{R}$.  

In the limit $q\dr 1$,  we recover the quantized cosine law for a flat triangle \cite{seth} expressed in terms of the quantized normals, modulo an overall sign and a factor $\demi$.
\bes\label{classical}
 \,^{(b)}\hat \tau^1 \cdot  \,^{(c)}\hat \tau^1 \, |\iota_{j_{a}j_{b}j_{c}}\ra &=& - \left( \frac{j_a(j_a+1)  - j_b(j_b+1)-j_c(j_c+1)}{\sqrt{ j_b(j_b+1)j_c(j_c+1)}} + \cO\left(\lambda^2\right)\right)\, |\iota_{j_{b}j_{c}j_{a}}\ra. 
\ees 
From the construction of the vector operators, in section \ref{vector operator}, we know that 
$$ \,^{(b)}\hat \tau^1 \cdot  \,^{(c)}\hat \tau^1  |\iota_{j_{a}j_{b}j_{c}}\ra = -\frac{2}{\sqrt{j_b(j_b+1)j_c(j_c+1)}}  \,^{(b)}\vec J \cdot  \,^{(c)}\vec J \,  |\iota_{j_{a}j_{b}j_{c}}\ra. $$
This allows to identify the source of the discrepancy  for the $\demi$ and the overall sign. In particular,  the global minus sign in \eqref{hyper} and \eqref{classical} with respect to the flat/hyperbolic cosine law comes simply from the definition of the scalar product we have used.

 Since $ \,^{(i)}\vec J $ is interpreted in the LQG formalism as the quantized normal to the edge of the triangle, in the deformed case, we interpret $\,^{(b)}\hat \bt^1$ and $\,^{(c)}\hat \bt^1$  as the quantized normals respectively of the edges $AC$ and $AB$, at the vertex $A$ of the hyperbolic triangle.


%

\medskip 

We can play with the normalization of the vector operators to have a better defined hyperbolic law. Indeed, we notice that  both \eqref{hyper} and \eqref{classical} diverge when $j=0$. Instead of taking the vector operator $\,^{(i)}\hat \bt^1$  with  normalization  $N^1_{j_i}=1$, we can consider  $\,^{(i)}\tilde \bt^1$ with normalization
\beq
\tilde N^1_j \equiv \f{\sqrt{\sinh (j\lambda)\, \sinh ((j+1)\lambda)}}{\sinh((j+\demi)\lambda)} \label{normalization 1}   \stackrel{q\dr 1}{\dr}  \frac{\sqrt{j(j+1)}}{j+\demi}. 
\eeq
In this case the cosine laws become well behaved for small $j$. 
\beq\label{hyperbis}
 \,^{(b)}\tilde\bt^1 \cdot  \,^{(c)}\tilde\bt^1 \, |\iota_{j_{b}j_{c}j_{a}}\ra  =  q \,    \frac{\cosh\frac{\ell_p}{2\ell_c} \cosh ((j_a+\demi) \lambda) - \cosh ((j_b+\demi)\lambda) \cosh((j_c+\demi)\lambda))}{{\sinh((j_b+\demi)\lambda)\, \sinh((j_c+\demi)\lambda)}} |\iota_{j_{b}j_{c}j_{a}}\ra.  
  \eeq

\medskip

When dealing with a non-zero cosmological constant and the Planck length, by dimensional analysis, one can expect to have a minimum angle  \cite{eugenio}.  This can now be explicitly checked. Setting $j_a=0$, we must have $j_b=j_c=j$ since we deal with an intertwiner, and the quantum cosine law \eqref{hyper} gives   
 \bes\label{nonzerocos}
 \theta_a^{min}(j)= \arccos \left(
q\frac{\cosh^2\frac{\lambda}{2} - \cosh^2 ((j+\demi)\lambda)}{{\sinh^2((j+\demi)\lambda) - \sinh^2\f\lambda2}}\right), 
\ees
which means that there is a non-zero minimum angle. When $\ell_p\dr0$ (classical limit) or $R\dr\infty$ (flat quantum limit), \eqref{nonzerocos} tends to 1, so we recover that the triangle is degenerated.  

\medskip

As expected, the angle observables can  be expressed in terms of the $\UQ$ generators.
\bes\nn
&& i>j, \,  \,^{(i)}\hat \bt^1 \cdot  \,^{(j)}\hat \bt^1 = \left( q^{-\f32} (- \cE_{ij}\cE_{ji} + \cE_{ii})+\f{1}{[2]}\cE_{ii}\cE_{jj} \right), \quad 
i<j, \,  \,^{(i)}\hat \bt^1 \cdot  \,^{(j)}\hat \bt^1=  \left( q^{\f12} ( -\cE_{ij}\cE_{ji} + \cE_{ii})+\f{q^2}{[2]}\cE_{ii}\cE_{jj} \right).
\ees

\medskip

\paragraph*{Length operator.}
The length operator is obtained by looking at the norm of the \textit{unnormalized}  vector operator $\,^{(i)} \bt^1$ with  normalization $N^i_j$. 
 \beq 
^{(i)}\bt^1\cdot ^{(i)}\bt^1| \iota_{j_bj_cj_a}\ra =\left(N^1_{j_i}\right)^2 | \iota_{j_bj_cj_a}\ra, \quad i=a,b,c.
 \eeq
 Keeping in mind that $\,^{(i)} \bt^1$ encodes the quantization of the normal, by inspection of the classical and quantum hyperbolic cosine law, it is natural to take 
\beq
N^1_j = \sqrt{\sinh^2((j+\demi)\lambda) - \sinh^2(\f\lambda2)} \textrm{ or }  \tilde {\bf N}^1_j = \sinh((j+\demi)\lambda).
\eeq  
 The  normalization  $ \tilde {\bf N}^1_j$ leads to the regularized hyperbolic cosine law \eqref{hyperbis}. We note therefore that the norm of the vector operator corresponds to a function of the length operator. The length is  quantized, with eigenvalue $(j+\demi)\ell_p$  as we have argued previously. The norm of the vector operator can be expressed in terms of the $\cE$ operators.
 \be
\,^{(i)} \bt^1 \cdot  \,^{(i)} \bt^1  =\frac{1}{[2]}  (-q\cE^2_i- (1+q\mone)\cE_i). 
 \ee
 
\medskip

\paragraph*{"Area" operator.}
In the flat case, one expresses the square of the area of the triangle in terms of a cosine and the norm of the normals so that the operator is easy to quantize, using  vector operators \cite{area-etera}.
 \beq
\cA^2= \f14\left(|\vec n_b|^2|\vec n_c|^2-(\vec n_c \cdot \vec  n_c)^2\right)
\eeq
We proceed in the same manner in the hyperbolic case. We do not consider the square of the area but instead the square of the sine of the area. Indeed, the area of an hyperbolic triangle is given in terms of the triangle angles \eqref{hyper area}. There are various ways to express functions of the area in terms of the edge lengths \cite{mend}. A convenient one will be 
\beq\label{mednykh}
\sin^2 \f{\cA}{2R^2}= \f{\sinh(\f{s}{2R})\sinh(\f{s-\ell_a}{2R})\sinh(\f{s-\ell_b}{2R})\sinh(\f{s-\ell_c}{2R})}{\cosh^2\f{\ell_a}{2R}\cosh^2\f{\ell_b}{2R}\cosh^2\f{\ell_c}{2R}},
\eeq
where $s=\demi(\ell_a+\ell_b+\ell_c)$. Of course, in the flat limit, $R\dr \infty$, we recover Heron's formula (see the above table). 


Playing with the cosine laws, we can express $\sin^2 \f{\cA}{2R^2}$ only in terms of the normals. 
\bes\label{areamend}
 \sin^2 \f{\cA}{2R^2}  = \f14\frac{ \sinh^2\f{\ell_b}{R}\sinh^2\f{\ell_c}{R}(1-\cos^2\theta_a) }{(\cosh^2\f{\ell_a}{2R}\cosh^2\f{\ell_b}{2R}\cosh^2\f{\ell_c}{2R})}= 2\frac{|\vec n_b|^2|\vec n_c|^2- (\vec n_b\cdot \vec n_c)^2}{(1+\sqrt{1+|\vec n_a|^2})(1+\sqrt{1+|\vec n_b|^2})(1+\sqrt{1+|\vec n_c|^2})}.
\ees
There is no difficulty in quantizing this expression since it only involves scalar products and norms of nromals, which upon quantization become operators that are diagonal and functions of the Casimir operator. There is therefore no ordering issue anywhere.  The area has also a discrete spectrum.

\section*{Outlook} \label{conclu}

\paragraph*{\bf Summary:}
Let us summarize the main results of our paper. We have recalled the definition of tensor operators for $\UQ$, with q real, which is the relevant case to study Euclidian 3d LQG with $\Lambda<0$ and Lorentzian 3+1 LQG with $\Lambda>0$. 

We have shown how they are the natural objects to construct observables for a $\UQ$ intertwiner. These operators are the key to study LQG defined in terms of  a quantum group as they provide sets of operators that transform well under the quantum group. We have generalized the $U(n)$ formalism to the quantum group $\UQ$. That is, we have shown how we can construct a closed algebra of observables (\ie invariant under $\UQ$) which can be related to the quantum group $\UUQn$. This means that the $\UQ$ intertwiner carries a $\UUQn$ representation, which we argued must be irreducible. 
We have constructed the natural generalization of the LQG geometric operators and interpreted them in the 3d Euclidian setting. We have shown that a three-leg $\UQ$ intertwiner  encodes the quantum state of an hyperbolic triangle. We have also shown how the presence of a cosmological constant leads to a notion of minimum angle as expected \cite{eugenio}. These results provide new evidences for the use of quantum group as a tool to encode the cosmological constant, in the LQG formalism. 

Note that the use of tensor operators can be also useful for dealing with lattice Yang-Mills theories built with $\UQ$ as gauge group. In particular it could be interesting to see how tensor operators can be useful to implement observables in the recent work \cite{bianca}. In fact, there are a number of interesting routes   open for exploration.

\medskip 

\paragraph*{\bf Hyperbolic polyhedra:}
We have studied the geometric operators in the context of 3d LQG. We have shown that they induce a quantum hyperbolic geometry. These operators should also be interpreted in the 3+1 LQG case. 
The vector operator acting on a leg $i$ would be interpreted as the quantization of the normal of the $i^{th}$ face of the polyhedron.  The  squared norm  of the vector operator acting on each leg would be now interpreted as a \textit{function} of the squared area  operator. This implies that in this case we still expect to have a discrete spectrum for the (squared) area. 
The angle operator would now encode the quantization of the dihedral angle, the angle between normals. One could then construct the analogue of the squared volume operator,  using the triple product between  vector operators. Following the intuition gained from looking at the area operator for the triangle, we would then expect to get an expression of a \textit{function} of the volume of the hyperbolic polyhedron. We leave for further investigations the properties of such operator, as well as other interesting geometric operators we could construct to probe the quantum geometry of hyperbolic polyhedra.

\medskip 

\paragraph*{\bf Phase space structure:}
One of our key results is that the quantum group spin networks can be used in the LQG context to introduce the cosmological constant.   Recent developments have shown that spin networks can be seen as quantum states of  flat discrete geometries, when $\Lambda=0$. The phase space structure is nicely described in the "twisted geometries" framework. Since we have identified the meaning of the quantum geometric operators,  built from the vector operators, this can provide some guiding lines in identifying the relevant phase space structure, \ie the notion of \textit{curved twisted geometries}. In particular, one knows that the classical analogue of a quantum group is a Poisson-Lie group, so we can expect to use this structure to define the curved twisted geometries. This is work in progress \cite{phase space}.

\medskip 

\paragraph*{\bf Other signatures and other signs for $\Lambda$:}
When defining tensor operators, we have focused on  $\UQ$ with $q$ real. This choice provided the relevant structure to study the physical case, 3+1 LQG with $\Lambda>0$.
However, there is a number of other cases to study. At the classical level, with $q=1$, we could explore the construction of tensor operators for $\SL(2,\R)$, which would be relevant for Lorentzian 2+1 LQG with $\Lambda=0$. Interestingly, the Wigner-Eckart theorem has not been defined   for $\SL(2,\R)$, that is there is no general formula for  tensor operators transforming as $\SL(2,\R)$ (non-unitary) finite dimensional and discrete representations\footnote{More precisely, there exists a definition of such tensor operators acting on the unitary (infinite dimensional) discrete  representation, provided by harmonic oscillators (Shwinger-Jordan trick). There is no such definition for operators acting on unitary (infinite dimensional) continuous representations. }. This is work in progress \cite{lorentzian}. It would be then relevant to discuss the quantum group version of this structure, which would be relevant for 2+1 Lorentzian gravity with $\Lambda\neq0$.  

Another interesting case to explore would  be $\UQ$ with $q$ root of unity, which would be relevant for 3d Euclidan LQG with $\Lambda>0$. We have not considered this case here as  $\UQ$ with $q$ root of unity is not a quasi-triangular Hopf algebra, but a quasi-Hopf algebra. This means that the construction in \cite{rittenberg} does not apply directly. On the other hand, the representation theory of 
$\UQ$ with $q$ root of unity can be trimmed of the unwanted features so that its recoupling theory can be well under control \cite{chari}. This is  why the Turaev Viro model can still be defined as it is.  It is then quite likely that we can  define the tensor operators in this case, in terms of their matrix elements, which would be  proportional to the Clebsh-Gordan coefficients. We leave this for further investigations.

\medskip

\paragraph*{\bf Hamiltonian constraint:}
LQG and spinfoams are supposed to be the two facets of the same theory. This can be shown explicitly only in the case $\Lambda=0$ case, in 3d \cite{ale-karim}. Recently,   an Hamiltonian constraint was constructed using the spinor formalism  \cite{valentin-etera}. It has been designed to encode a recursion relation on the $6j$ symbol and hence by construction, it relates the Ponzano-Regge model to the LQG approach. Now that we have generalized the spinor approach to the quantum group case, we can construct a $q$-deformed version of this Hamiltonian constraint. It would essentially encode the recursion relation of the $q$-deformed $6j$ symbol. Hence this new $q$-deformed Hamiltonian constraint  would relate the Turaev-Viro model and  LQG with a cosmological constant. This is work in progress \cite{hamiltonian}.  

\bigskip

\textbf{Aknowledgements:}
We would like to thank A. Baratin, V. Bonzom, B. Dittrich, L. Freidel, E. Livine for many interesting discussions and comments.

 \appendix

\section{Hyperbolic cosine law}\label{cosine law} 
  
 Consider the upper 2d  hyperboloid $H_+^2$, embedded   into $\R^3$, with curvature  $-R^{-2}=\Lambda$, where $R$ is the radius of  curvature.   
\bes
H_+^2&=& \lbrace \vec x\in \R^3, \; x_1>0, \, x_i \eta^{ij}x_j= x_1^2-x_2^2-x_3^2 = |\vec x |^2= R^2 \rbrace.
\ees
On $H_+^2$, consider three points $A,B,C$ and the  geodesics joining them: we obtain an hyperbolic triangle. Without loss of generality, we can always assume that $A$ sits at the origin of $H_+^2$, that is as a point of $\R^3$, it is given by the vector $\vec A = (R, 0,0)$. The points $B$ and $C$ are then obtained from $\vec A$ by performing a boost $L_c, \, L_b$ with respective rapidity $c$ and $b$.  Explicitly, 
\bes \label{boostrot}
&& \vec B = L_c \vec A, \quad \vec C = L_b \vec A. 
\ees
As a consequence, we have $\la \vec  A, \vec A\ra = |\vec A|^2= |\vec B|^2=|\vec C|^2=R^2$. 

\medskip
Consider the normalized space-like vectors $\hat u_{AB}, \, \hat  u_{AC}\in T_AH_+^2$, the tangent plane of $H_+^2$ at the point $A$. They are the tangent vectors to the geodesics joining respectively $A$ to $B$ and $A$ to $C$. By construction, these vectors are orthogonal to $\vec A$. 
\bes
\hat  u_{AB}=  \frac{\vec B - \frac{1}{R^2}\langle\vec A,\vec B\rangle \vec A}{\left|\vec B - \frac{1}{R^2}\langle\vec A,\vec B\rangle \vec A\right| }, &\quad& 
\hat  u_{AC}=  \frac{\vec C - \frac{1}{R^2}\langle \vec A,\vec C\rangle\vec A}{\left| \vec C - \frac{1}{R^2}\langle\vec A,\vec C\rangle \vec A\right| }
\label{u3}
\label{u}
\ees
Since we are dealing with an homogeneous space, we express the lengths $\ell_i$ of the geodesic arcs   using the dimensionful parameter $R$, such that $\ell_{c}=R c$, $\ell_{b}=R b$ as well as $\ell_{a}=R a$.   


\medskip

By definition, we know that the angle between two geodesics which intersect is defined in terms of the angle between the tangent vectors. If we focus in particular on the angle $\alpha$ between the arcs $AB$ and $AC$, we have 
\beq
\cos \alpha = \langle  \hat  u_{AB},  \hat  u_{AC} \rangle.
\eeq
Using the expression of the tangent vectors, we obtain the hyperbolic cosine law.
\bes
\cos \alpha  &= & \frac{-\cosh \frac{\ell_a}{R} + \cosh \frac{\ell_b}{R} \cosh \frac{\ell_c}{R}}{\sinh \frac{\ell_b}{R}\sinh \frac{\ell_c}{R}}
\label{hypcos}\ees
In the flat case,  performing the limit $R\dr\infty$ in  \eqref{hypcos}, we recover the al Khashi rule
\beq \label{alkashi}
\cos \alpha = \frac{- \ell_a^2+ (\ell_b^2 + \ell_c^2)}{2\ell_b\ell_c}.
\eeq

\section{Useful formulae}\label{WCGcoef}\label{useful}\label{CG app} \label{6j app}\label{rmat app}
These formulae are taken from the book \cite{BiedenharnBook}.

\paragraph{\qCG}
An explicit expression of the \qCG in the van der Waerden form is given as
\bes
\CG{c@{}c@{}c}{j_1& j_2& j\\ m_1 & m_2 & m}&:= &\delta_{m, m_1+m_2} q^{\f{1}{4}(j_1+j_2-j)(j_1+j_2+j+1)+\f{1}{2}(j_1m_2-j_2m_1)}\Delta(j_1j_2j) \\
& \times & \left( [j_1+m_1]! [j_1-m_1]! [j_2+m_2]! [j_2-m_2]! [j+m]! [j-m]! [2j+1]\right)^{\f12} \\
& \times & \sum_n \f{(-1)^n q^{-\f{n}{2}(j_1+j_2+j+1)}}{[n]![j_1+j_2-j-n]![j_1-m_1-n]![j_2+m_2-n]![j-j_2+m_1+n]![j-j_1-m_2+n]!}.
\ees
where the triangle function $\Delta$ is given by
\be
\Delta(abc):= \left(\f{[a+b-c]![a-b+c]![-a+b+c]!}{[a+b+c+1]!} \right)^{\f12}.
\ee 
 For $q \rightarrow 1$ the \qCG coefficients  reduce to the usual CG coefficients in the van der Waerden form.
 
 \medskip
The   \qCG coefficients have two orthogonality relations. 
\bes
\sum_{m_{1},m_{2}} \CGa \CG{c@{}c@{}c}{j_1& j_2& j'\\m_1 & m_2 & m'} &=& \delta_{jj'}\delta_{mm'} \label{CGortho1}\\ 
\sum_{j,m} \CGa \CG{c@{}c@{}c}{j_1& j_2& j\\m'_1 & m'_2 & m}&=& \delta_{m_{1}m'_{1}}\delta_{m_{2}m'_{2}}.\label{CGortho2}
\ees 
 Note that in the first equation, we have assumed that $j_{1}, j_{2}$ and $j$ satisfy the triangle conditions.
 
 \medskip
 The \qCG coefficients  have some symmetries. We list the most relevant ones for our concerns.
\bes \label{recurrence}
\CGa &=& (-1)^{j_{1}+j_{2}-j}\mCG{c@{}c@{}c}{j_1& j_2& j\\-m_1 &- m_2 &- m} \label{CGsym1}\\ 
\CGa& =&  (-1)^{j_{1}+j_{2}-j}\mCG{c@{}c@{}c}{j_2& j_1& j\\m_2 &m_1 & m}  \label{CGsym2}\\
\CGa &= & (-1)^{j-j_{2}-m_{1}}q^{\frac{m_{1}}{2}}\sqrt{\f{[2j+1]}{[2j_{2}+1]}}\CG{c@{}c@{}c}{j_1& j& j_{2}\\-m_1 & m & m_{2}}.  \label{CGsym3}
\ees  
 
 \medskip
 
 The value of some specific CG coefficients.
 
\be\label{WCG00}
\CG{c@{}c@{}c}{j_1& j_2 & 0\\m_1& m_2 & 0}= \delta_{j_1, j_2} \, \delta_{m_1, -m_2} \, \f{(-1)^{j_1-m_1}q^{\f{m_1}{2}}}{\sqrt{[2j_1+1]}}.
\ee
 
\bes
&& \CG{c@{}c@{}c}{1& 1 & 1\\ 1 & 0 & 1}= q^{1/2}\sqrt{\f{[2]}{[4]}}, \quad  \CG{c@{}c@{}c}{1& 1 & 1\\ 0 & 1 & 1}=-q^{-1/2}\sqrt{\f{[2]}{[4]}}, 
\quad  \CG{c@{}c@{}c}{1& 1 & 1\\ -1 & 0 & -1}= -q^{-1/2}\sqrt{\f{[2]}{[4]}}, \quad   \CG{c@{}c@{}c}{1& 1 & 1\\ 0 & -1 & -1}=q^{1/2}\sqrt{\f{[2]}{[4]}}, \nn\\
&& \CG{c@{}c@{}c}{1& 1 & 1\\ 1 & -1 & 0}=\sqrt{\f{[2]}{[4]}}, \quad \CG{c@{}c@{}c}{1& 1 & 1\\ -1 & 1 & 0}=-\sqrt{\f{[2]}{[4]}},  \quad \CG{c@{}c@{}c}{1& 1 & 1\\ 0 & 0 & 0}=\sqrt{\f{[2]}{[4]}}\left(q^{\demi} - q^{-\demi}\right).
\ees

\paragraph{$q$-$6j$-symbol}
The \qsj is invariant under the rescaling $q\dr q \mone$. It satisfies the following orthogonality relation.
 \bes
\sum_{j}  \qsixj{ccc}{b&c&j\\k&a& n}   \qsixj{ccc}{a&b&m\\c&k& j }&=&\delta_{mn}\label{6jortho1}
\ees
The contraction of two  \qsj can give another one. This is a useful property for us.
\bes\nn
\sum_{m} (-1)^{a+b+c+k-j-m-n}q^{\demi(a(a+1)+b(b+1)+c(c+1)+k(k+1)-j(j+1)- m(m+1)-n(n+1))}   \qsixj{ccc}{a&b&m\\c&k& j}  \qsixj{ccc}{a&c&n\\k&b& m}&=&  \qsixj{ccc}{ a&c&n\\b&k& j}   \label{6jcomb}
 \ees
 It has some symmetries when moving some of its elements.
 \bes
 \qsixj{ccc}{ a&b&m\\c&k& j }=    \qsixj{ccc}{ c&k&m\\a&b& j} \label{6jsym}
 \ees 
 A specific value of the \qsj which is relevant to us is 
 \bes
 \qsixj{ccc}{ j_{1}&j_{1}&1\\j_{2}&j_{2}& j_{3} }&=& (-1)^{j_{1}+j_{2}+j_{3}}\frac{[j_{2}+j_{3}-j_{1}][j_{1}+j_{3}-j_{2}]- [j_{1}+j_{2}-j_{3}][j_{1}+j_{2}+j_{3}+2]}{\left([2j_{1}][2j_{1}+1][2j_{1}+2][2j_{2}][2j_{2}+1][2j_{2}+2]\right)^{\demi}}.
 \ees

 \be\label{CG1}
\CG{c@{}c@{}c}{\f12 & \f12 & 1\\ \f12 & \f12 & 1}=1 = \,\CG{c@{}c@{}c}{\f12 & \f12 & 1\\ -\f12 & -\f12 & -1}, \quad \CG{c@{}c@{}c}{\f12 &\f12 & 1\\ \f12 & -\f12 & 0}=\f{q^{-\f14}}{\sqrt{[2]}}, \quad \CG{c@{}c@{}c}{\f12 &\f12 & 1 \\ -\f12 & \f12 & 0}=\f{q^{\f14}}{\sqrt{[2]}},
\ee
 
\paragraph{$\cR$-matrix and deformed permutation}
The $\cR$-matrix for $\UQ$ can be expressed in terms of the \qCG. 

\bes\label{R1}
\left(\cR^{j_1j_2}\right) ^{m_1 m_2}_{m'_1 m'_2} &=& \sum_{j,m} q^{-\demi \left( j_{1}(j_{1}+1)+ j_2 (j_2+1)- j(j+1) \right)}  \CG{ccc}{j_{1}&j_{2 } & j\\ m_{1}& m_{2} & m}\; \mCG{ccc}{j_{1}&j_{2 } & j\\ m'_{1}& m'_{2} & m} \\
 &=& \sum_{j,m}(-1)^{j_{1}+j_{2}-j} \,q^{-\demi \left( j_{1}(j_{1}+1)+ j_{2}(j_{2}+1)- j(j+1) \right)} \CG{ccc}{j_{1}&j_{2 } & j\\ m_{1}& m_{2} & m} \;\CG{ccc}{j_{2}&j_{1 } & j\\ m'_{2}& m'_{1} & m},\label{R2}
\ees
with $m_{1}+m_{2}= m'_{1}+m'_{2}$ and $m'_{1}-m_{1}\geq 0$ (this is zero otherwise). The second equation has been obtained using the symmetries of the \qCG coefficients.

The inverse of the $\cR$-matrix is obtained from the above formulae by setting $q\dr q\mone$. 
\bes\label{Rmone}
\left({\cR^{-1}}^{j_1j_2}\right) ^{m_1 m_2}_{m'_1 m'_2}   &=& \sum_{j,m}(-1)^{j_{1}+j_{2}-j} \, q^{\demi \left( j_{1}(j_{1}+1)+ j_{2}(j_{2}+1)- j(j+1) \right)} \mCG{ccc}{j_{1}&j_{2 } & j\\ m_{1}& m_{2} & m} \;\mCG{ccc}{j_{2}&j_{1 } & j\\ m'_{2}& m'_{1} & m}\\
  &=& \sum_{j,m}(-1)^{j_{1}+j_{2}-j}\, q^{\demi \left( j_{1}(j_{1}+1)+ j_{2}(j_{2}+1)- j(j+1) \right)} \CG{ccc}{j_{2}&j_{1 } & j\\ m_{2}& m_{1} & m} \;\CG{ccc}{j_{1}&j_{2 } & j\\ m'_{1}& m'_{2} & m}
\ees
One can check that this is true by evaluating $\cR\mone \cR$ and use the orthogonality properties of the \qCG coefficients. Furthermore we can check that when $q\dr 1$, we recover that the $\cR$-matrix is simply the identity map (for this one uses the classical version of \eqref{CGsym2} and the orthogonality relation \eqref{CGortho2}).

\medskip

We are interested in the deformed permutation $\psi_\cR = \psi \cR$ (resp. $\psi_\cR\mone = \cR\mone \psi$), which means that instead of considering ${\cR}^{j_1j_2}$ (resp. ${\cR^{-1}}^{j_1j_2}$), we consider  ${\cR}^{j_2j_1}$ (resp. ${\cR^{-1}}^{j_2j_1}$). The relevant formula for ${\cR}^{j_2j_1}$ is obtained from \eqref{R2} by exchanging $j_{1}$ and $j_{2}$.



\end{document}